\documentstyle[
aps,prb]{revtex}

\begin{document}
\draft
\title{Intra--chain correlation functions and shapes of homopolymers
with different architectures in dilute solution}
\author{Edward G.~Timoshenko\thanks{Corresponding author. 
Internet: http://darkstar.ucd.ie; 
E-mail: Edward.Timoshenko@ucd.ie}
}
\address{
Theory and Computation Group,
Department of Chemistry, University College Dublin,
Belfield, Dublin 4, Ireland}

\author{Yuri A.~Kuznetsov}
\address{
Centre for High Performance Computing Applications, 
University College Dublin,
Belfield, Dublin 4, Ireland}

\author{
Ronan Connolly
}
\address{
Theory and Computation Group,
Department of Chemistry, University College Dublin,
Belfield, Dublin 4, Ireland}

\date{\today}
\maketitle

\begin{abstract}
We present results of Monte Carlo study of the monomer--monomer
correlation functions, static structure factor
and asphericity characteristics of a single homopolymer in the 
coil and globular states for three distinct architectures of the chain: 
ring, open and star.
To rationalise the results we introduce the dimensionless 
correlation functions rescaled via
the corresponding mean--squared distances between monomers.
For flexible chains with some architectures these functions 
exhibit a large degree of universality by falling onto a single 
or several distinct master curves.
In the repulsive regime, where a stretched exponential times
a power law form (de Cloizeaux scaling) can be applied, the corresponding 
exponents $\delta$ and $\theta$ have been obtained. 
The exponent $\delta=1/\nu$ is found to be universal for flexible
strongly repulsive coils and in
agreement with the theoretical prediction from improved 
higher--order Borel--resummed renormalisation group calculations.
The short--distance exponents $\theta_{\upsilon}$ of an open 
flexible chain are in a good agreement with the theoretical 
predictions in the strongly repulsive regime also.
However, increasing the Kuhn length
in relation to the monomer size leads to their fast cross--over towards the
Gaussian behaviour. Likewise, a strong sensitivity of various exponents
$\theta_{ij}$ on the stiffness of the chain, or on the number of arms in
star polymers, is observed. The correlation functions in 
the globular state are found to have a more complicated oscillating
behaviour and their degree of universality  has been reviewed.
Average shapes of the polymers in terms of the asphericity characteristics,
as well as the universal behaviour in the static structure factors, 
have been also investigated.
\end{abstract}

\vskip 1cm
\pacs{PACS numbers: 36.20.-r, 36.20.Ey, 61.25.Hq}

\section{Introduction}
\label{sec:intro}

Knowledge of the two--point correlation functions $g^{(2)}_{ij}(r)$
of a polymer chain
provides one with profound statistical information on the internal structure 
of the conformation \cite{CloizeauxBook} as well as a direct way for
computing numerous pair--wise observables of interest: 
mean energy ${\cal E}$, mean--squared distances between monomers
${\cal D}_{ij}$, mean--squared radius of gyration ${\cal R}_g^2$,
static structure factor $S(p)$, and many others. While some of these 
observables permit direct experimental measurement and simple evaluation 
at the level of mean--field theories, the whole range of information
contained in the pair correlation functions is only accessible in part to
the most sophisticated of modern polymer theories
\cite{CloizeauxBook,Eu,Taylor}.  On the other hand, computer simulations
based on Monte Carlo Metropolis technique, if performed on a large
enough scale for gathering sufficiently large statistics, can yield 
accurate values of $g^{(2)}_{ij}(r_i)$ on a given spherical 
mesh $r_i$ via computing histograms of monomer
occurrences in spherical layers around other monomers.

Considerable simulational effort in this area
\cite{Freire,Valleau,Bishop94,Bishop95,Baumgartner}
has been largely inspired by the predictions of
the universal scaling behaviour in the field theoretical approaches
(e.g. based on the de Gennes' $n\to 0$ formalism) developed by
J.~de Cloizeaux \cite{CloizeauxBook,desCloizeaux80},
B.~Duplantier \cite{Duplantier,DuplantierLet} and some others 
\cite{Oono,Schafer}.
As simulations have grown larger, the agreement with the theoretically 
computed exponents has improved to a satisfactory level \cite{Freire,Valleau}.
Many results, however, were only known for an open homopolymer chain in
the repulsive regime (good solvent), apart from the exact 
case of the Gaussian chain. 

In the current work we shall undertake a more systematic study
of the pair correlation functions and related observables
for polymers of varying length in a wide range of the degree of
polymerisation $N=100-700$ units;
with either open, ring or star (with different
numbers of arms $f=3,6,9,12$) architectures; with 
flexible or semi--flexible (with stiffness $\lambda=1,5$) chain;
as well as for the cases of good and poor solvents. 

Generally, the introduction of a more complex chain architecture,
stiffness, or attraction, would pose a serious challenge for 
field--theoretic formulations and renormalisation group calculations,
and hence only a few scaling results are known for the correlation
functions in such cases.
Moreover, the SO(n)-$\varphi^4$ field theory, which relates to the
polymer diagrams, was 
previously only studied up to several loops \cite{CloizeauxBook,Schafer}. 
Recently, however, new results up to seven loops have been reported in 
$d=3$ dimensions \cite{Guida}. These, by themselves, would not necessarily
yield a better precision because of poor asymptotic convergence
of the series, but aided by the Borel--resummation
renormalisation group techniques, the convergence can be dramatically
improved and errors in the exponents reduced. Thus, it is important
to re--examine the agreement of the theory with simulations even in the
previously studied case of an open flexible coil, as well as
to investigate other architectures. Bearing in mind that
the theoretical scalings only apply for sufficiently long chains,
it is also instructive to analyse the data for different chain lengths, 
and to study dependencies on various model parameters generally to address
the degree of universality in the scaling laws.

\section{Model}
\label{sec:model}

The current coarse--grained homopolymer model is based
on the following Hamiltonian (energy functional) \cite{CombStar,Torus,CopStar} 
in terms of the monomer coordinates, ${\bf X}_i$:
\begin{eqnarray}
H & = & \frac{k_B T}{2\ell^2}  \sum_{i\sim j} \kappa_{ij}
        ({\bf X}_i - {\bf X}_{j})^2
      + \frac{k_B T }{2\ell^2} \sum_{i\approx j \approx k} \lambda_{ijk}
        ({\bf X}_{i} + {\bf X}_{k} - 2{\bf X}_j)^2  \nonumber \\
  & & + \frac{1}{2} \sum_{ij,\ i\not= j} V (|{\bf X}_i - {\bf X}_j|)
\equiv \frac{1}{2} \sum_{ij,\ i\not= j} H_{ij} (|{\bf X}_i - {\bf X}_j|).
\label{cmc:hamil}
\end{eqnarray}
Here the first term represents the connectivity structure of the polymer with
harmonic springs of a given strength $\kappa_{ij}$ introduced between
any pair of connected monomers (which is denoted by $i\sim j$).
The second term represents the bending energy penalty given by the
square of the local curvature with a characteristic stiffness $\lambda_{ijk}$
between any three consecutively connected monomers (which is
denoted by $i\approx j \approx k$) in the form of the Kratky--Harris--Hearst
term \cite{KratkyHarris}.
Finally, the third term represents pair--wise non--bonded interactions
between monomers such as the van der Waals forces and so on.
In a simple homopolymer model we can adopt the Lennard--Jones of the potential,
\begin{equation}\label{VLJ}
V(r) = \left\{
\begin{array}{ll}
+\infty, &  r < d \\
V_0 \left( \left( \frac{d}{r}\right)^{12}
- \left( \frac{d}{r} \right)^{6} \right), &  r > d
\end{array}
\right.,
\end{equation}
where there is also a hard core part with the monomer diameter $d$
(below we choose $d=\ell$ without any lack of generality). 

We use the Monte Carlo technique with the standard Metropolis 
algorithm \cite{AllenTild}, which converges to the Gibbs equilibrium ensemble,
based upon the implementation 
described by us in Ref. \onlinecite{Torus}.  Simulations were performed for an 
ensemble of $250$ (or $50-100$ when additional symmetries of the system
were present) of independently generated initial conditions,
each of which has been first equilibrated by typically $20\,N^3$ of attempted
Monte Carlo steps. To ensure proper equilibration, the long-- and short--time
behaviour of the global observables, such as e.g. 
the ${\cal E}$ and ${\cal R}_g^2$,
was monitored, as these cease to have 
a global drift and start exhibiting characteristic
fluctuating behaviour around well defined mean values upon reaching
the equilibrium state for given values of the interaction parameters.
After reaching the equilibrium, $1000$ of statistical measurements
for each initial condition have been performed. These measurements were
separated by typically $40\, N^2$ of attempted Monte Carlo
steps in between to ensure statistical independence of sampling.
Thus, overall the averaging statistics was of order $Q=10^5$ in all cases.
The mean value and error of sampling of an observable  $A$ are then
given by the arithmetic mean 
$\langle A \rangle=(1/Q)\sum_{\gamma}^Q A_{\gamma}$
and by $\pm \sqrt{(\Delta A)^2/Q}$ respectively.
For more details on our Monte Carlo  technique we refer to the
previous papers Refs. \onlinecite{Torus,CopStar}.

\section{Definitions}
\label{sec:defs}

The intra--chain correlation function of a pair of monomers $i$ and $j$ 
is defined as,
\begin{equation}
g^{(2)}_{ij}({\bf r}) \equiv \biggl\langle \delta({\bf X}_i 
- {\bf X}_{j}-{\bf r}) \biggr\rangle = \frac{1}{4\pi r^2} \biggl\langle 
\delta(|{\bf X}_i - {\bf X}_{j}|-r) \biggr\rangle.
\end{equation}
The second equation establishes that it is a function of radius $r=|{\bf r}|$
only due to spatial isotropy (SO(3) rotational symmetry).
We may note that
this function should, strictly speaking, be named distribution function,
but since $g^{(2)}_{ij}(r)\to 0$ when $r\to \infty$ 
because of the chain connectivity, we apply the term `correlation function'  to
$g^{(2)}_{ij}(r)$ itself rather than to the quantity 
$g^{(2)}_{ij}(r)/(g^{(1)})^2-1$, which would vanish 
as $r\to \infty$  in the case of simple liquids.
The function is normalised to unity via:
$\int d^3{\bf r}\, g^{(2)}_{ij}({\bf r})=1$.
Note that the correlation functions exactly satisfy the excluded
volume condition,
\begin{equation}
\label{cond}
g_{ij}^{(2)}(r)=0 \qquad \mbox{for}\ r < d,
\end{equation}
due to the choice of the hard--core part in the non--bonded potential
Eq. (\ref{VLJ}).

The mean energy and the static structure factor (SSF) then can be computed
as follows,
\begin{eqnarray}
{\cal E}& =& \langle H \rangle = \frac{1}{2} \sum_{ij,\ i\not= j}
             \int d^3{\bf r}\, H_{ij}(|{\bf r}|)\, g^{(2)}_{ij}({\bf r}), \\
S(p) &=& \frac{1}{N}\sum_{ij} \tilde{g}^{(2)}(|{\bf p}|), \qquad
\tilde{g}^{(2)}({\bf p}) = \left\langle \exp(i{\bf p}\,({\bf X}_i -
    {\bf X}_j) \right\rangle = \frac{1}{2\pi^2}
    \int_0^{\infty}r^2\,dr\,\frac{\sin(pr)}{pr}\,g^{(2)}_{ij}(r),
\end{eqnarray}
where tilde indicates the 3-d Fourier transform.
The mean--squared distance between monomers $i$ and $j$ is,
\begin{equation}\label{Dd}
{\cal D}_{ij} \equiv \langle D_{ij} \rangle =  \biggl\langle
({\bf X}_i - {\bf X}_{j})^2 \biggr\rangle = \int d^3{\bf r}\,|{\bf r}|^2\, 
g^{(2)}_{ij}({\bf r}),
\end{equation}
which we defined here without the traditional factor of $1/3$ as compared
to previous paper Ref.~\onlinecite{Torus}. Then the mean--squared 
radius of gyration is simply,
\begin{equation}\label{Rg}
{\cal R}_g^2 = \langle R_g^2 \rangle = \frac{1}{2N^2}
               \sum_{ij,\ i\not= j} {\cal D}_{ij}.
\end{equation}

Finally, for analysis of the polymer shape we can define the
non--averaged {\it shape tensor}
\cite{Solc,Diehl,Rudnick,Jagodzinski,Aronovitz},
\begin{equation}
\label{tens}
Q^{\alpha \beta}({\bf X}_i) = \frac{1}{2\,N^2} \sum_{ij,\ i\not= j}
(X_i^{\alpha}- X_j^{\alpha})(X_i^{\beta}- X_j^{\beta}),
\end{equation}
which is related to the inertia tensor $I^{\alpha\beta}=
\delta^{\alpha\beta}\mbox{tr}\, Q-Q^{\alpha\beta}$ in the case when all monomers
have equal masses. Let us also introduce the traceless tensor 
$\hat{Q}^{\alpha\beta}=Q^{\alpha\beta}-\frac{1}{3}
\delta^{\alpha\beta}\mbox{tr}\, Q$
and the `mean' value of the eigenvalues $\bar{q}\equiv (1/3)\mbox{tr}\,Q$,
where the trace of $Q$ according to Eqs. (\ref{Rg},\ref{tens})
coincides with the non--averaged squared radius of gyration, namely:
$\mbox{tr}\,Q= \sum_{a=1,2,3}q^{(a)}=R_g^2$.
One can study the mean ratios of the eigenvalues \cite{Solc,Jagodzinski},
\begin{equation}
\label{La}
\lambda^{(a)} = \left\langle\frac{q^{(a)}}{\mbox{tr}\,Q}\right\rangle 
=\left\langle\frac{q^{(a)}}{3\bar{q}}\right\rangle, \qquad
\sum_{a=1,2,3}\lambda^{(a)}=1.
\end{equation}
In addition, we can define the following two asphericity characteristics, 
\cite{Solc,Diehl,Rudnick,Jagodzinski,Aronovitz},
\begin{eqnarray}
{\cal A}_3= \langle A_3 \rangle & = & \left\langle\frac{3}{2}
            \frac{\mbox{tr}\,\hat{Q}^2}{(\mbox{tr}\,Q)^2} 
\right\rangle= \left\langle
\frac{1}{6}\sum_{a=1,2,3}\frac{(q^{(a)}-\bar{q})^2}{\bar{q}^2}\right\rangle ,
          \qquad 0\ \mbox{(sphere)} \leq {\cal A}_3 \leq 1
          \ \mbox{(collinear)}, \label{Aa}\\
{\cal S}_3 = \langle S_3 \rangle & = & \left\langle\frac{27\det
          \hat{Q}}{(\mbox{tr}\,Q)^3} \right\rangle
         = \left\langle \frac{\prod_{a=1,2,3}(q^{(a)}-\bar{q})}{\bar{q}^3}
         \right\rangle, \qquad -\frac{1}{4} \ \mbox{(oblate)} \leq {\cal S}_3
         \leq 2 \ \mbox{(prolate)}. \label{Ss}
\end{eqnarray}
Alternatively, we can consider modified quantities, 
\cite{Diehl,Rudnick,Aronovitz}
\begin{equation}\label{hatAS}
\hat{{\cal A}}_3 \equiv \frac{\langle \bar{q}^2 A_3\rangle}{\langle \bar{q}^2
\rangle}, \qquad 
\hat{{\cal S}}_3 \equiv \frac{\langle \bar{q}^3 S_3\rangle}{\langle \bar{q}^3
\rangle}, 
\end{equation}
which are more amenable to analytical treatments, although less
sensitive on the shape of conformations.

\section{Correlation functions}\label{sec:res}

\subsection{Flexible ring polymers}

Ring polymers are the simplest objects of study since they
possess exact translational invariance along the chain,
so that all single--point observables are constants
and all pair--wise observables are functions of the separation
in the chain indices $n=|i-j|$ rather than of both indices \cite{Torus},
i.e. ${\cal D}_{ij}={\cal D}_{0n}$, $g^{(2)}_{ij}(r) = g^{(2)}_{0n}(r)$.

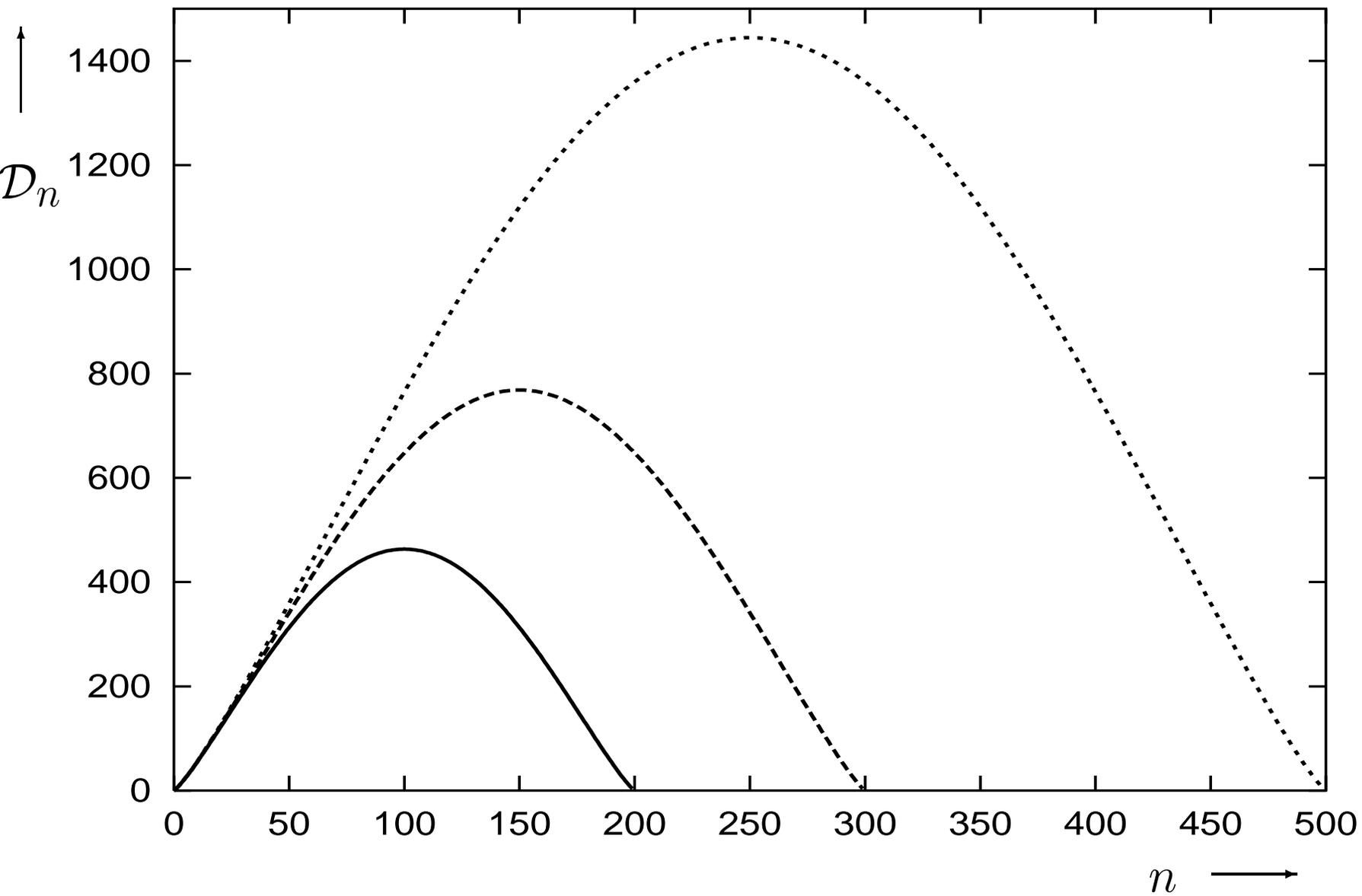
\begin{figure}
\caption{ \label{fig:DkRing}
Plot of the mean--squared distances ${\cal D}_n$ (in units of $\ell^2$) 
of flexible ($\lambda=0$) homopolymer rings
in a good solvent ($V_0=0$) vs the chain index $n$.
Curves correspond to the values of the degree of polymerisation
$N=200,\ 300,\ 500$ (from bottom to top).
}
\end{figure}

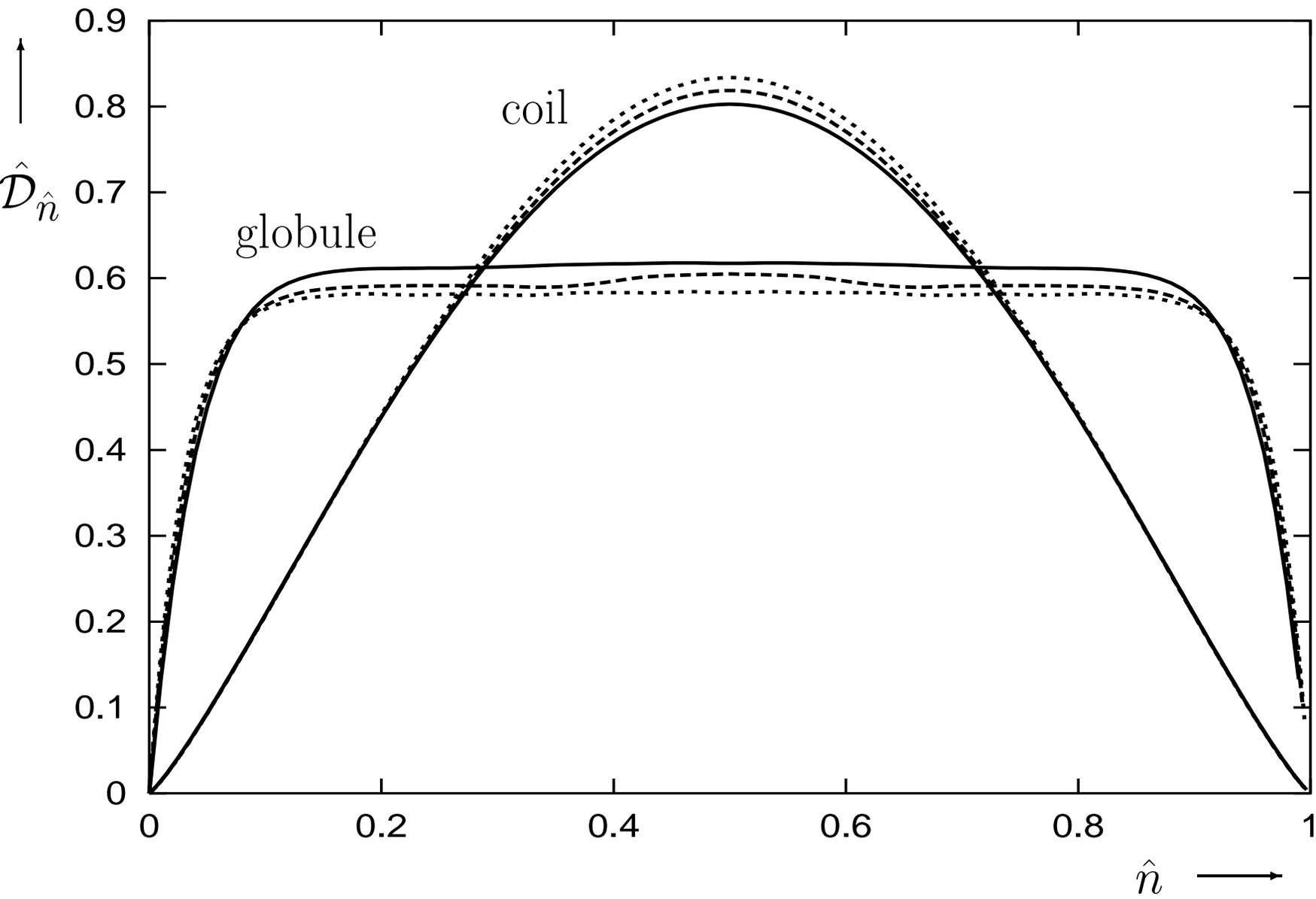
\begin{figure}
\caption{ \label{fig:DkResc}
Plot of the rescaled mean--squared distances
$\hat{{\cal D}}_{\hat{n}}=N^{-2\nu}\,
{\cal D}_n$ (in units of $\ell^2$) vs the rescaled chain index $\hat{n}=n/N$
of flexible homopolymer rings
in a good solvent ($V_0=0$, $\nu=\nu_F$): bell--shaped curves 
for $N=200$ (solid line), $N=300$ (long--dashed line), 
$N=500$ (short--dashed line);
and in a poor solvent ($V_0=6\,k_B T$, $\nu=1/3$): trapezoid--shaped curves
for $N=100$ (solid line), $N=150$ (long--dashed line), 
$N=200$ (short--dashed line).
}
\end{figure}

In Fig. \ref{fig:DkRing} we exhibit the mean--squared distances ${\cal D}_n$
of flexible ring homopolymers vs the separation $n$. 
There is an additional symmetry property
of rings such that ${\cal D}_n$ is also symmetric with respect to the middle
(i.e. maximal separation along the chain), namely ${\cal D}_n={\cal D}_{N-n}$.
This function has a characteristic bell--shape (which would become
somewhat deformed in case of semi--flexible polymers), and this shape
agrees very well with that from the Gaussian Self--Consistent
(GSC) method (see e.g. Fig. 2a in Ref. \onlinecite{Torus}). This function
also appears to possess an approximate scaling transformation property, so that
curves with different $N$ can be superimposed into each other by
the transformations: $\hat{n} = n/N$, $\hat{{\cal D}}_{\hat{n}}= 
{\cal D}_n/N^{2\nu_F}$,
where $\nu_F\approx 3/5$ is the value of the Flory swelling exponent.
Such rescaled quantities $\hat{{\cal D}}_{\hat{n}}$ 
are shown in Fig. \ref{fig:DkResc} for both the coil (in a good solvent)
and the globule (in a poor solvent). For the coil, $\hat{{\cal D}}_{\hat{n}}$
coincide with each other very 
closely everywhere except around the middle of the range
$|0.5 - \hat{n}| \lesssim 0.2$, where such agreement is less accurate.
However, without rescaling, ${\cal D}_n$ is strongly dependent on both $n$
and $N$. The dependence on $N$ is relatively weak only for $n$ corresponding
to a few polymer links, where connectivity is directly manifested.

Similarly, without rescaling, the intra--chain pair correlation 
functions $g^{(2)}_n(r)$ are strongly dependent on both $n$ and $N$. 
However, if we introduce the rescaled pair correlation function in terms
of the dimensionless variables,
\begin{equation}
\hat{g}^{(2)}_{ij}(\hat{r}) \equiv {\cal D}_{ij}^{3/2}\,\, 
g^{(2)}_{ij}\left(r\right),
\qquad \hat{r} \equiv \frac{r}{{\cal D}_{ij}^{1/2}},
\end{equation}
these would permit much more straightforward comparison with each other.
Note that $\hat{g}^{(2)}_{ij}(\hat{r})$ satisfies the following two
normalisation conditions:
\begin{equation}
\label{normtwo}
\int_0^{\infty} d\hat{r}\,\hat{r}^2\,\hat{g}^{(2)}_{ij}(\hat{r})=
\int_0^{\infty} d\hat{r}\,\hat{r}^4\,\hat{g}^{(2)}_{ij}(\hat{r})= \frac{1}{4\pi}.
\end{equation}

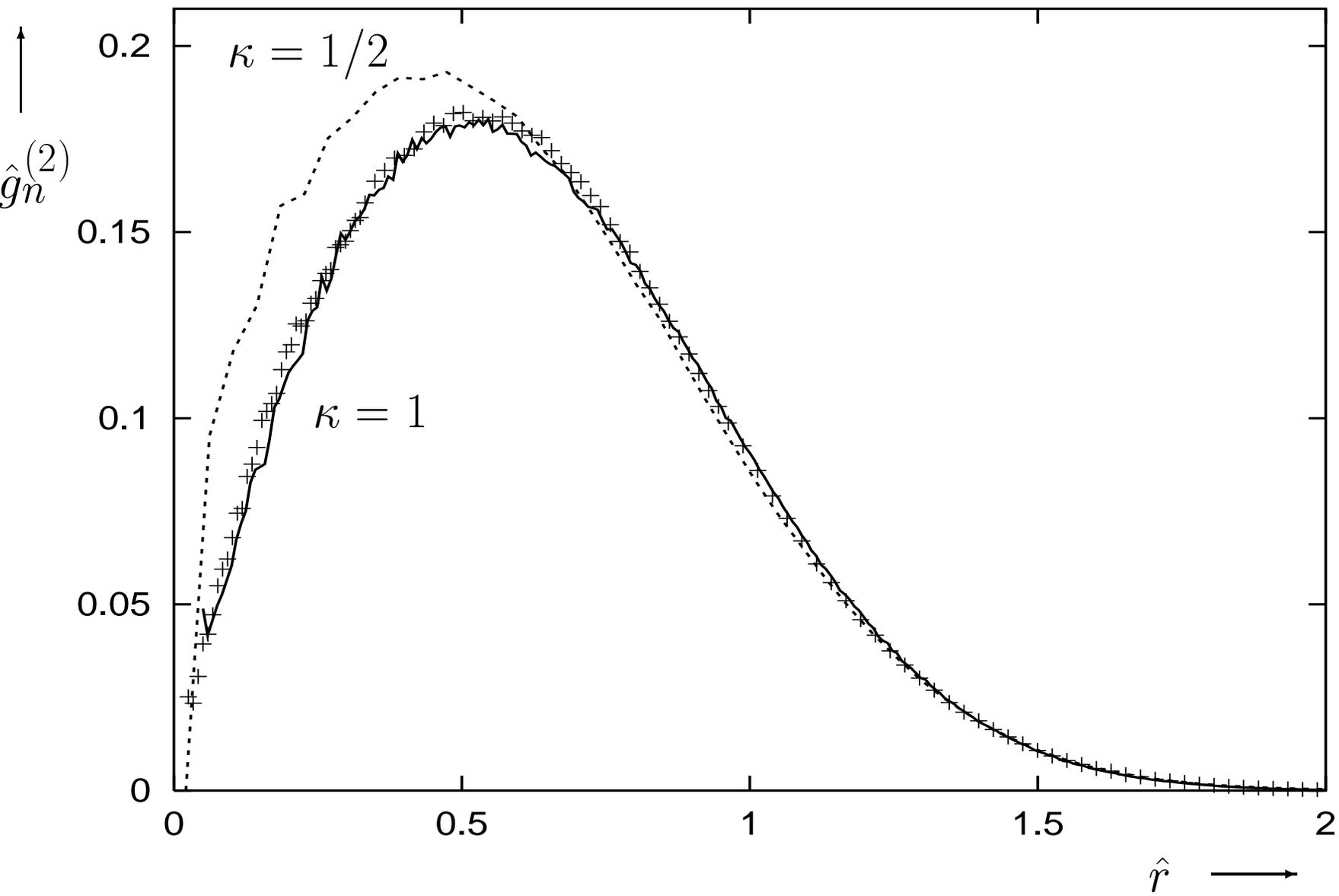
\begin{figure}
\caption{ \label{fig:gkRing}
Plot of the rescaled correlation function $\hat{g}^{(2)}_n(\hat{r})$ 
of homopolymer rings in a good solvent ($V_0=0$) vs the 
rescaled spatial separation $\hat{r}$ corresponding to the chain index
of half--ring ($n=N/2$).
Curves correspond to the values of the degree of polymerisation
$N=200$ (solid line) and $N=700$ (pluses).
Dashed line corresponds to half--ring $n=N/2$ of $N=200$ homopolymer,
but with twice weaker connectivity constants $\kappa_{ij}=1/2$.
}
\end{figure}

Thus, in Fig. \ref{fig:gkRing} we present $\hat{g}^{(2)}_{n}(\hat{r})$
for $n$ corresponding to the half--ring chain separation for different
values of the degree of polymerisation $N$. 
It is important to emphasise that the dashed curve here
corresponds to a different connectivity constant and 
this will be discussed later on.
The two lower curves coincide with each
other remarkably well, except for very small values of $\hat{r}$, which are
affected by the hard--core part of volume interactions in the area of
direct steric repulsion $r\simeq d$ (see also Eq. (\ref{cond}). 
The same is true of 
$\hat{g}^{(2)}_{n}(\hat{r})$ curves corresponding to different chain
separations $n$, except very small values of $n \lesssim n^{\dagger} 
\simeq 5$ 
for which ${\cal D}_n$ is of order of several $d^2$ units, i.e. where 
${\cal D}_n d^2\gtrsim 1$.
We may remark that at such very small $n$ any particular polymer model
(e.g. bead--and--anharmonic--springs one or a freely--jointed one)
would in any case exhibit its model--specific features, and hence 
would no longer produce a universal behaviour.

One can see that the function $\hat{g}^{(2)}(\hat{r})$ is non--monotonic:
it increases for small $\hat{r}$, reaches a maximum
at around $\hat{r} \simeq 1/2$
and then decreases in a Gaussian--like manner, being almost zero for
$\hat{r}\gtrsim 2$. The short--distance behaviour, which is often called the
{\it correlation hole} \cite{Schafer}, is an effect of the excluded
volume interactions coming from repulsions mediated through
higher than direct binary contacts described by Eq. (\ref{cond}).

From this interpretation one would expect that the correlation hole size
should decrease upon decreasing the excluded volume size $d$ relative
to the Kuhn length $\ell/\sqrt{\kappa}$. This is indeed the
case as can be seen
from the dashed curve in Fig. \ref{fig:gkRing}, which has been obtained
for a polymer with a twice weaker spring constant $\kappa$. The correlation
hole size decreases, and the value of $\hat{g}^{(2)}$ there correspondingly
increases to keep the normalisation condition and hence
the function reaches the maximum earlier. 
The effect is the opposite at distances $\hat{r}\simeq 1$,
but much weaker, and the tail at larger $\hat{r}$ is
practically unaffected by this change of connectivity parameter.
We would like to emphasise that while the change in the correlation
hole with the Kuhn length is quite dramatic, it is by no means
unexpected. Evidently, as much weaker springs are considered (with $d$ fixed), 
we should approach the Gaussian law for the correlation 
function, $\hat{g}^{(2)}_n \propto
\exp(-3\,\hat{r}^2/2)$, as well as for the mean--squared
distances, and the effect of cross--over behaviour is indeed so clearly seen
in Fig. \ref{fig:gkRing}.

\subsection{Globule of a ring chain}

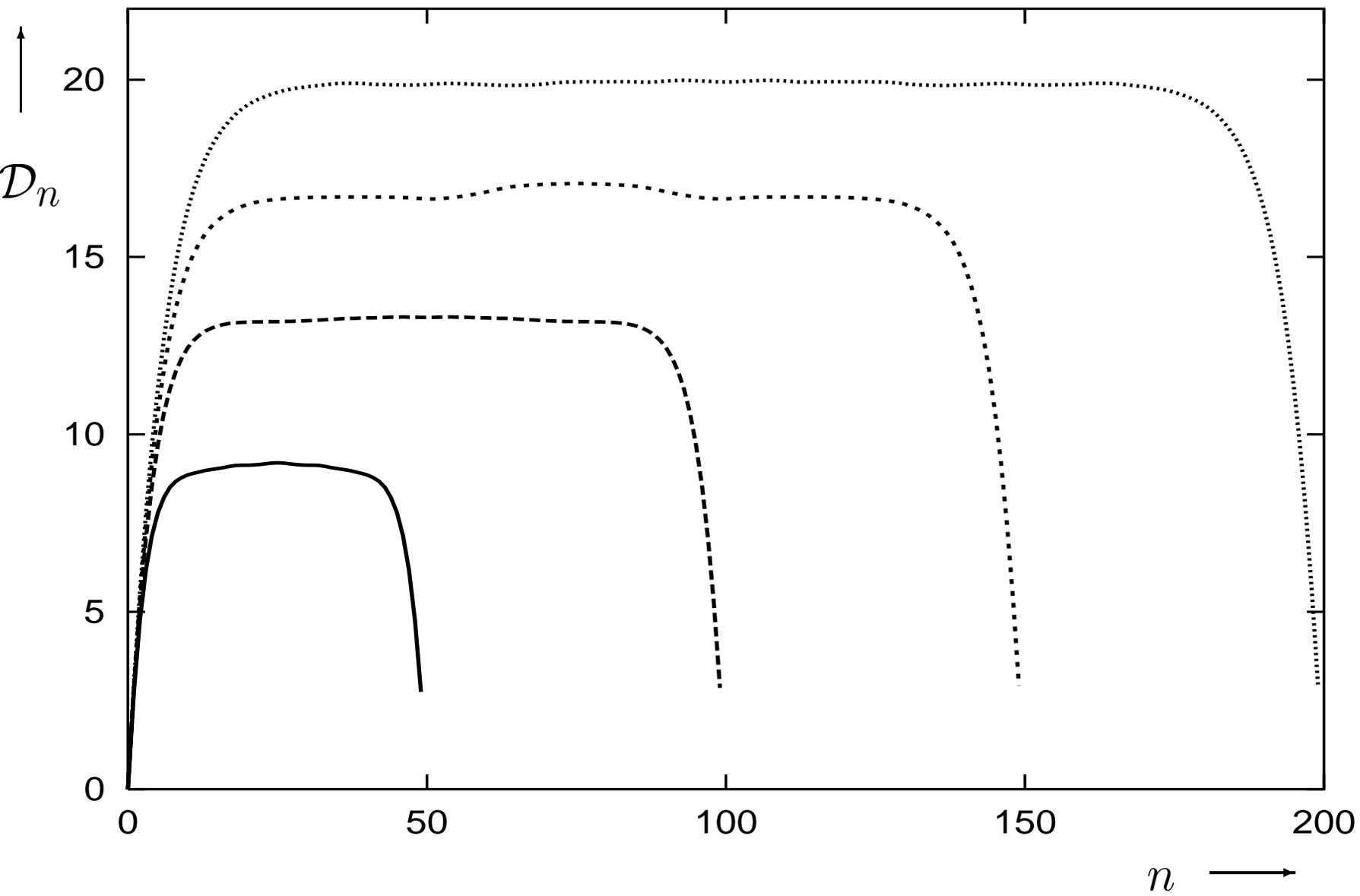
\begin{figure}
\caption{ \label{fig:DkGlob}
Plot of the mean--squared distances ${\cal D}_n$ (in units of $\ell^2$) 
of homopolymer rings
in a poor solvent ($V_0=6\,k_B T$) vs the chain index $n$.
Curves correspond to the values of the degree of polymerisation
$N=50,\ 100,\ 150, \ 200$ (from bottom to top).
}
\end{figure}

After collapse of the coil, the mean--squared distances 
${\cal D}_n$ of the globule at the van der Waals 
attraction $V_0=6\,k_B T$ are
shown in Fig. \ref{fig:DkGlob} vs the chain separation $n$ for different
chain lengths $N$. These functions increase rapidly at small $n$,
quickly reaching a stable plateau scaling as ${\cal D}_n \sim N^{2/3}$
at $n\sim n^{*}\sim N^{2/3}$.
The plateau is related to the nearly constant density within the globule.
The shapes of ${\cal D}_n$ in Fig. \ref{fig:DkGlob} agree with those
from the GSC method (see
solid line in Fig. 2a in Ref. \onlinecite{Torus}
and curve marked LLG in Fig. 2 in Ref. \onlinecite{QzMess}). 

Rescaled curves $\hat{{\cal D}}_{\hat{n}} = N^{-2/3}\,
{\cal D}_{\hat{n}}$ vs $\hat{n}=n/N$ in Fig. \ref{fig:DkResc}
lie fairly close to each other. With increasing
$N$ the plateau height in $\hat{{\cal D}}_{\hat{n}}$ slightly
decreases, reaching its asymptotic
value as the cross--over scale vanishes: $\hat{n}^{*}\sim N^{-1/3}\to 0$,
so that the normalisation $\sum_{\hat{n}<1} \hat{{\cal D}}_{\hat{n}}=
{\cal R}_g^2\,N^{-2/3}$ remains constant.

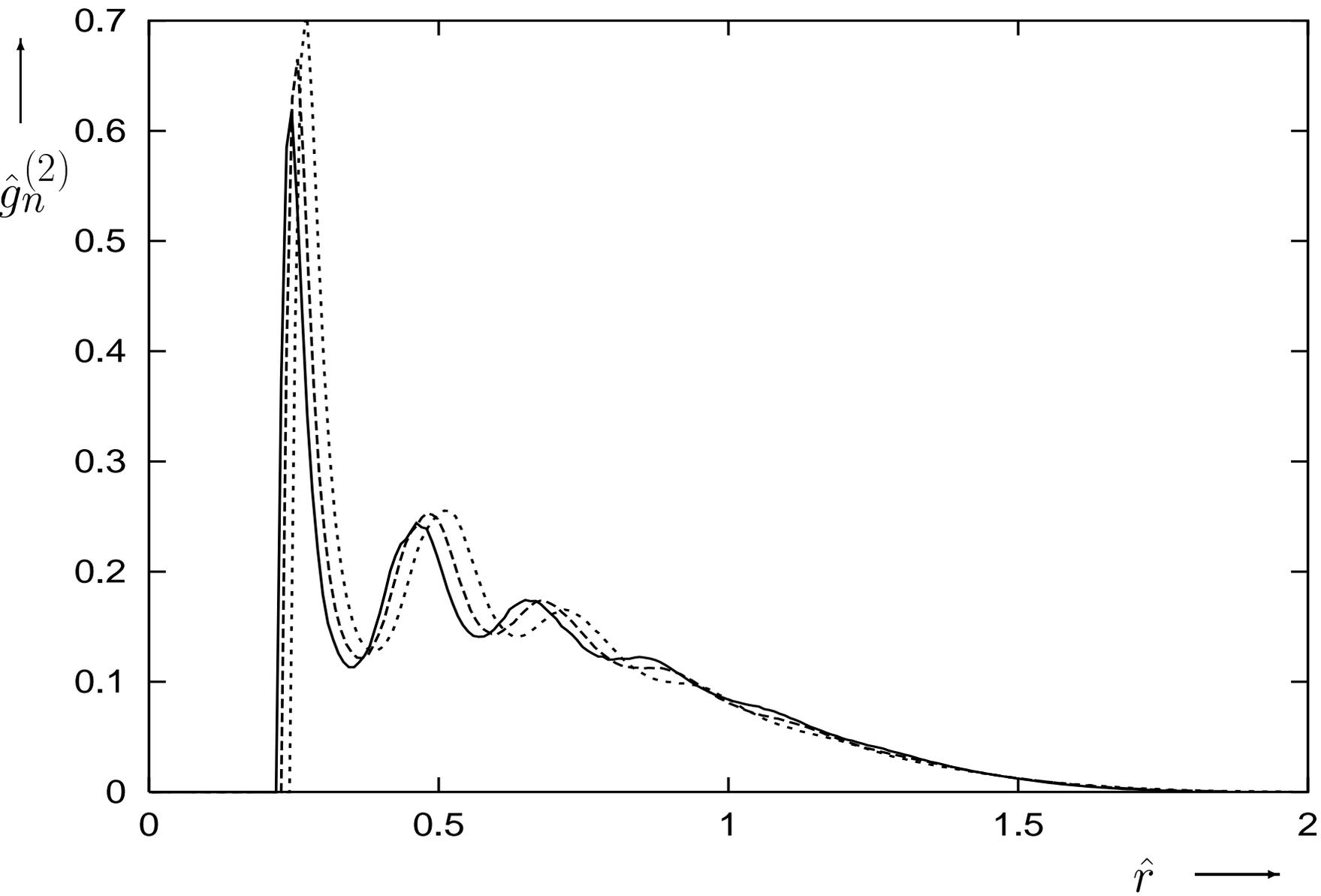
\begin{figure}
\caption{ \label{fig:gkGlob}
Plot of the rescaled correlation function $\hat{g}^{(2)}_n(\hat{r})$ 
of homopolymer rings in a poor solvent ($V_0=6\,k_B T$) vs the 
rescaled spatial separation $\hat{r}$ for the degree of polymerisation
$N=200$. 
Curves correspond to the values of the chain index
$n=10$ (short--dashed line), $n=15$ (long--dashed line) and $n=100$ (solid 
line).
}
\end{figure}

In Fig. \ref{fig:gkGlob} we present
the rescaled correlation functions $\hat{g}_{n}^{(2)}(\hat{r})$  of the globule
for several different values of $n$. These, for small $\hat{r}$, 
have several peaks of increasing width and decreasing height
located at integer multiples of the excluded volume diameter $r=d\,n$,
which is similar to the behaviour of correlation functions in simple
liquids \cite{HansenMacDonald}.
For large $\hat{r}$ this behaviour changes into a monotonically decreasing
tail. All functions $\hat{g}_{n}^{(2)}(\hat{r})$ 
for chain separations $|N/2 - n| \gtrsim n^*$,
where $n^*$ is the same cross--over index as in ${\cal D}_n$, 
lie very close to each other and to the half--ring function 
$\hat{g}_{N/2}^{(2)}(\hat{r})$. Thus, as long as the chain separations $n$
are not directly affected by the chain connectivity, the idea of
a single master function in terms of the dimensionless variables
is still valid for the globule, although the function is much 
more complicated than for the coil.
This is consistent with the
fact that peaks in $g^{(2)}_n(r)$ occur at $r=d\,n$, so that
for $\hat{g}^{(2)}_n(\hat{r})$ to fall onto a single function
${\cal D}_n$ has to become $n$-independent,
which indeed happens beyond $n^*$.

Importantly, the curves $\hat{g}^{(2)}_n$ do not fall onto a
single master curve for different chain lengths $N$ here.
Indeed, the number of peaks in $\hat{g}^{(2)}_n$ 
can be estimated as proportional to the linear size of the
globule, which is of order $\propto N^{1/3}$. Thus, for a very large
$N$ we shall have a function containing a sharp and narrow first peak
followed by many oscillations decreasing in amplitude around a constant
level (related to the constancy of the density inside the globule) and
ended by a monotonically decreasing tail.

\subsection{Open chain}

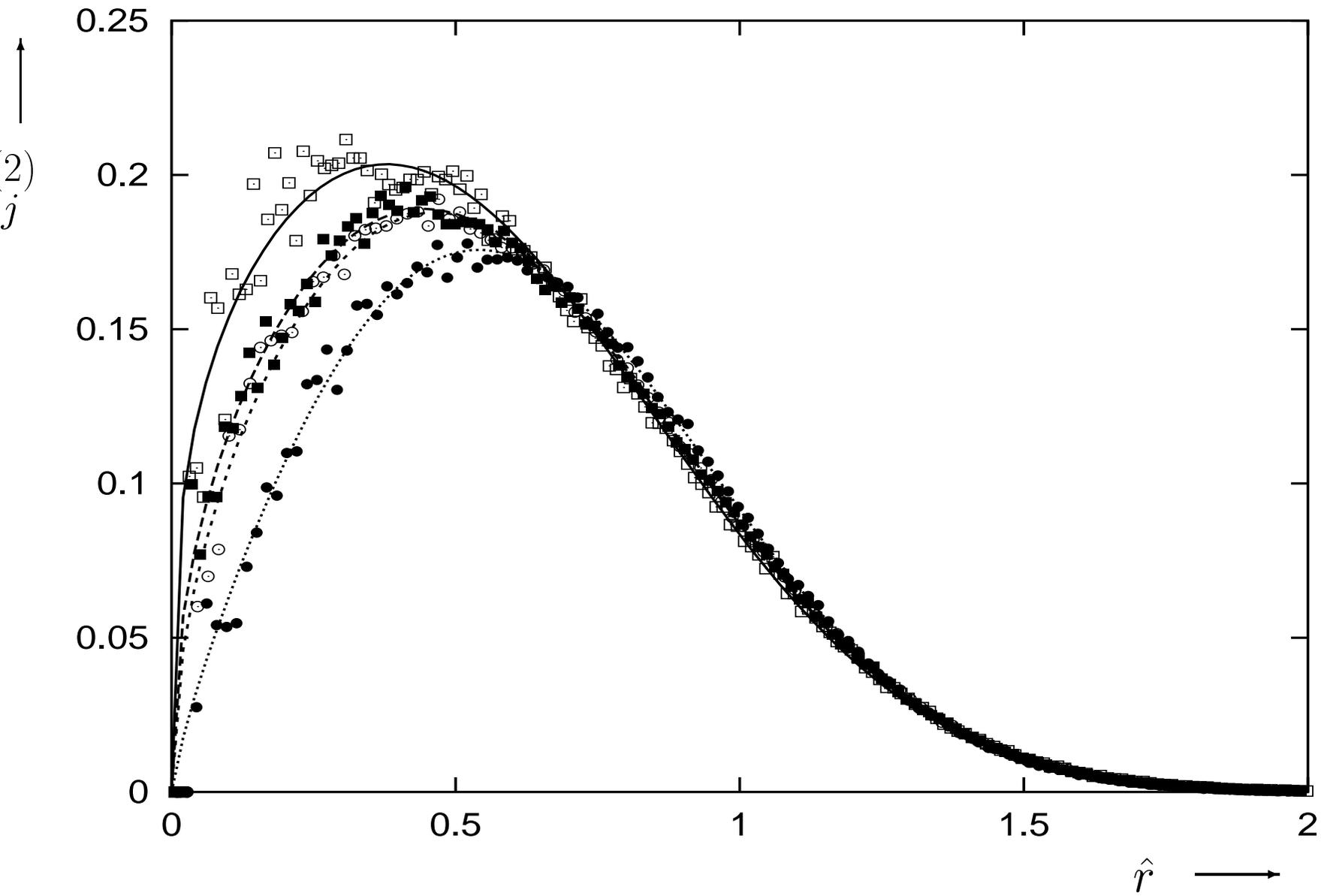
\begin{figure}
\caption{ \label{fig:gkOpen}
Plot of the rescaled correlation function $\hat{g}^{(2)}_n(\hat{r})$ 
of an open homopolymer in a good solvent ($V_0=0$) vs the 
rescaled spatial separation $\hat{r}$ for the degree of polymerisation
$N=200$. 
Curves correspond to the following values of the chain indices:
end--end $i,j=1,200$ (solid line and empty quadrangles), 
end--three quarters $i,j=1,150$ (long--dashed line and filled quadrangles), 
end--middle $i,j=1,100$ (short--dashed line and empty circles), 
and quarter--three quarters $i,j=50,150$ (dotted line and filled circles) 
from top to bottom in the left of the figure.
Here lines have been obtained by fitting via Eq. (\ref{FitA}) and points
correspond only to a subset of actual data to make the figure less
cluttered.
}
\end{figure}

Rescaled correlation functions of an open chain are presented in Fig.
\ref{fig:gkOpen}. The ring symmetry no longer applies here, so that
$\hat{g}_{ij}^{(2)}(\hat{r})$ now explicitly depends on both chain indices.
We note that the reflection symmetry $g_{N+1-i,N+1-j}^{(2)}=g_{ij}^{(2)}$
is the only exact symmetry of an open chain.
Thus, in obtaining the correlation functions, we may average over
two possibilities for the end--internal and internal--internal
correlations. As for the end--end correlation function
only single possibility exists hence
leading to $\sqrt{2}$ times worse statistics in this case.
Since the averaging statistics is the worst for an open chain as compared
to other architectures considered in this paper,
we present the correlation functions here via data points and 
smooth curves approximating them (this will be discussed in more
detail in Sec. \ref{sec:scaling}).

Clearly, the end monomers have a special r\^{o}le due to entropic reasons,
but correlations of internal monomers on large separations within a very
long chain should behave in a similar manner to those of an equivalent 
ring. Indeed, the function $\hat{g}_{N/4,3N/4}^{(2)}(\hat{r})$ (denoted by
dotted line in Fig. \ref{fig:gkOpen}) practically coincides with that of
a ring (solid line in Fig. \ref{fig:gkRing}).

However, the functions involving the end monomers have distinct shapes.
Short--dashed curve of $\hat{g}_{1,N/2}^{(2)}(\hat{r})$ has a higher
peak and a somewhat smaller correlation hole area than the middle--middle
correlation function; the trend continues for the long--dashed curve of
$\hat{g}_{1,3N/4}^{(2)}(\hat{r})$, which shows a weak
cross--over behaviour; and the end--end correlation 
function $\hat{g}_{1,N}^{(2)}(\hat{r})$ (solid line) has the tallest
peak and the narrowest correlation hole area. Interestingly, for
$\hat{r} \gtrsim 1.5$ all functions still have practically identical 
monotonically decreasing tails.

\subsection{Semi--flexible ring}\label{subsec:stiff}

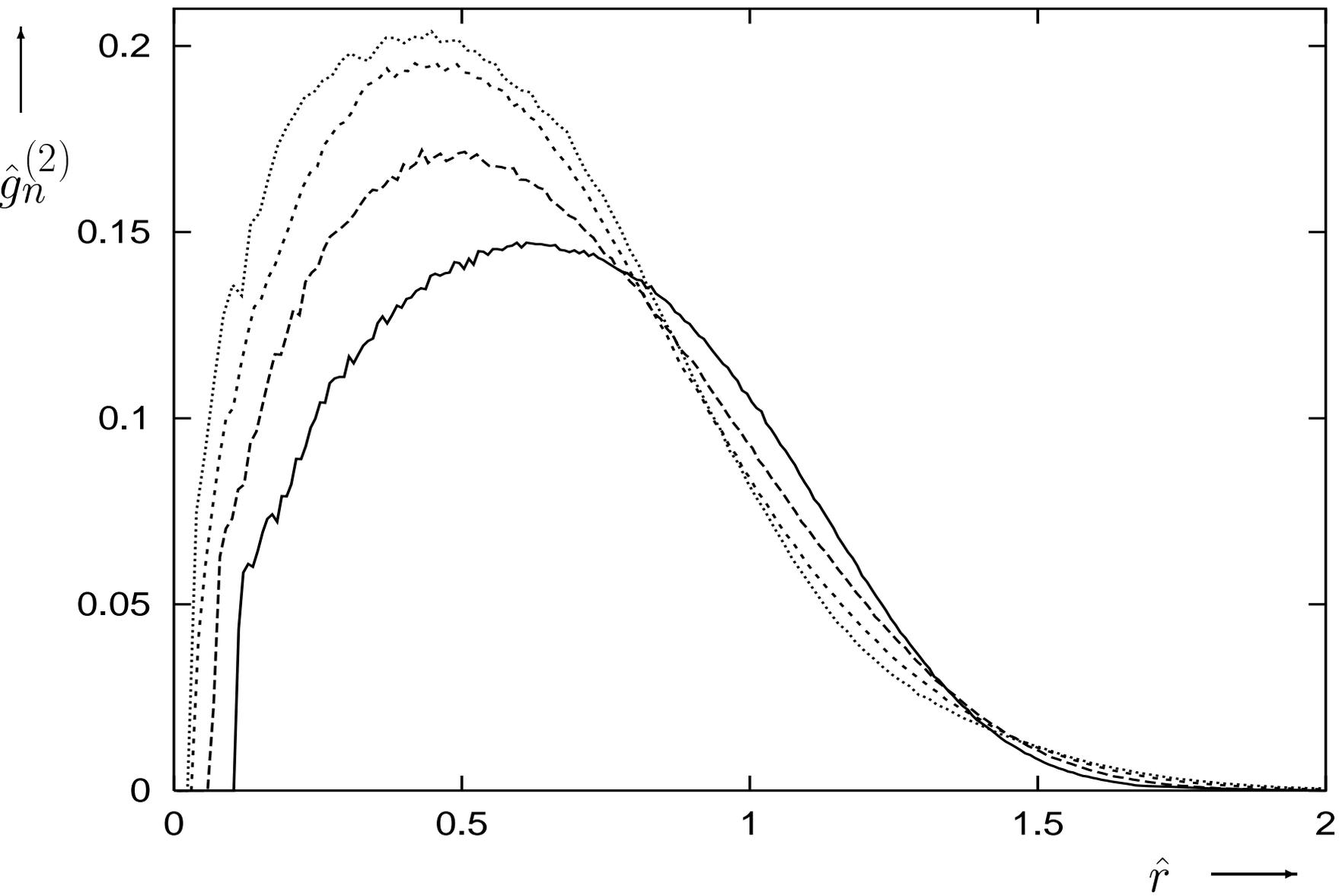
\begin{figure}
\caption{ \label{fig:gkStiff}
Plot of the rescaled correlation function $\hat{g}^{(2)}_n(\hat{r})$ 
of a semi--flexible ($\lambda/\kappa=1$) homopolymer ring in a good 
solvent ($V_0=0$) vs the 
rescaled spatial separation $\hat{r}$ for the degree of polymerisation
$N=200$.
Curves correspond to the values of the chain index
$n=10$ (solid line), $n=20$ (long--dashed line) and $n=50$ (short--dashed line),
and $n=100$ (dotted line).
Note also that these curves run 
from bottom to top in the left of the figure.
}
\end{figure}

Now then, it would be interesting to understand the effect of chain stiffness
determined by the $\lambda_{ijk}$ constants in Eq. (\ref{cmc:hamil}).
Thus, let us consider a semi--flexible ring polymer of $N=200$ units 
with $\lambda_{ijk}=1$ for each triplet of consequently 
connected monomers and zero otherwise \cite{Torus}.
In Fig. \ref{fig:gkStiff} we present $\hat{g}^{(2)}_n(\hat{r})$ functions for
different values of chain separation $n$. As opposed to similar functions
of a flexible ring ($\lambda=0$) in Fig. \ref{fig:gkRing}, in this case
$\hat{g}^{(2)}_n(\hat{r})$ no longer fall onto a single master curve.

A striking feature for relatively small $n$ is the expansion of
the correlation hole area at small $\hat{r}$. The latter is now determined
not only by the excluded volume interactions but also by the tendency 
of the chain to remain locally straight via the stiffness term.
The lowest curve for $n=10$ in Fig. \ref{fig:gkStiff} shows that this
effect is still present for $n$ considerably larger than the persistence
length of the chain, which is of order $\lambda$.
Upon increasing $n$ the correlation hole shrinks,
producing a steeper growth at small $\hat{r}$, and for large enough $n$
the hole becomes even smaller than that of the flexible ring.
This may be understood as follows. Due to the stiffness effect neighbouring
monomers along the chain are being pushed away from the current monomer,
thus providing an easier access for contacts with
it for monomers which are more distant along the chain.
The mismatch of $\hat{g}^{(2)}_n$ for different $n$
is present also at intermediate values of $\hat{r}$, where a complicated
cross--over occurs, and, unusually, even the tails of the distributions
do not match perfectly here.
Thus, stiffness completely removes the universality of rescaled
correlation functions which was present in the flexible case for topologically
equivalent monomers. Moreover, manifestations of stiffness are also
strongly $N$-dependent \cite{Torus} and the consideration of the limit
$N\to\infty$ is highly non--trivial as it would require 
renormalisation of $\lambda$, which might
significantly change the physical persistent length of the chain.

\subsection{Star polymers}

Now let us turn our attention to the case of flexible homopolymer stars
with arm length $(N-1)/f=50$. We refer the reader for notations, further
details and previous results on stars 
to our preceding work in Ref. \onlinecite{CopStar}.

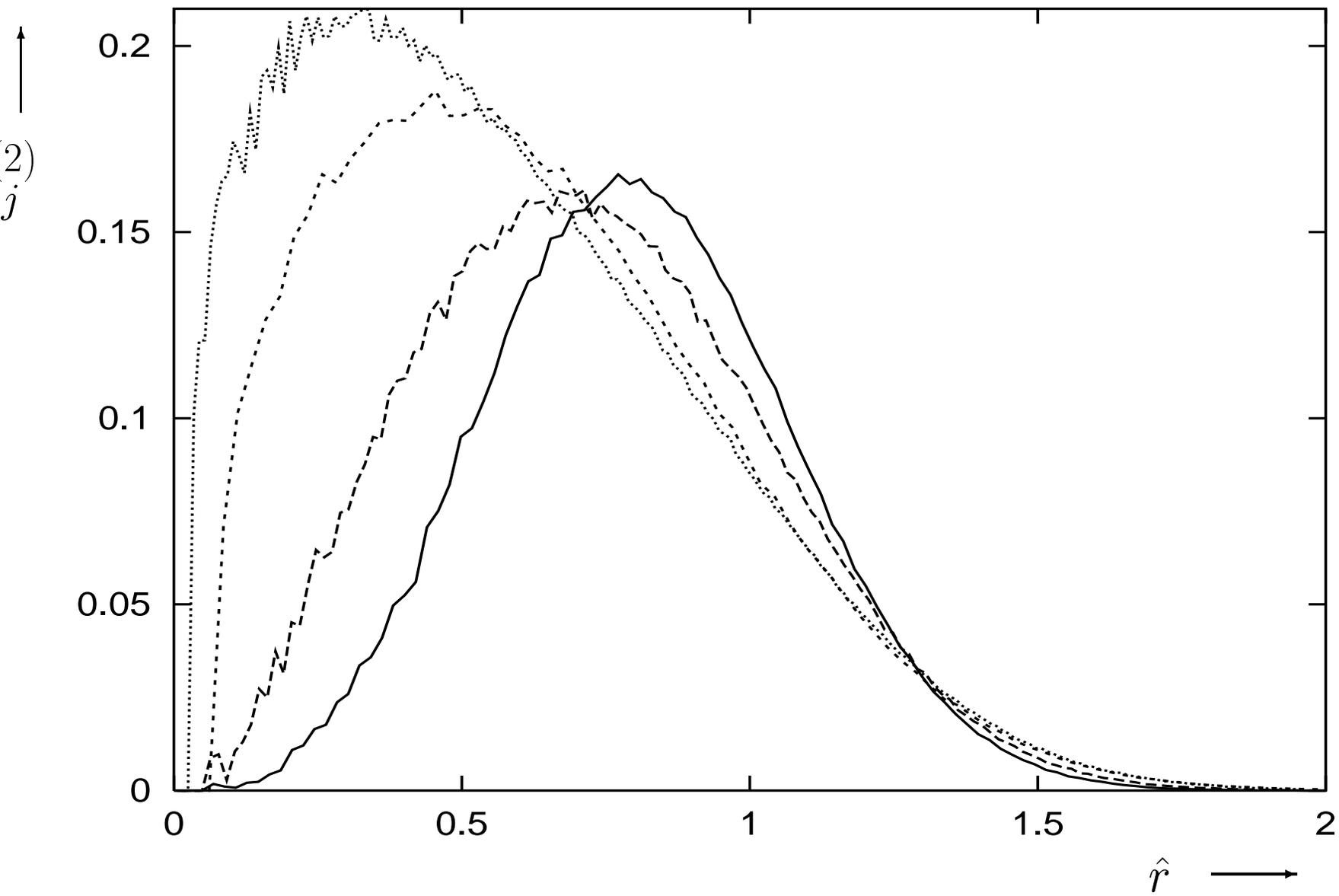
\begin{figure}
\caption{ \label{fig:gkStar}
Plot of the rescaled correlation functions $\hat{g}^{(2)}_{ij}(\hat{r})$ 
of a homopolymer star with the number of arms $f=12$ and arm length
$(N-1)/f=50$ in a good solvent ($V_0=0$) vs the 
rescaled spatial separation $\hat{r}$.
Curves correspond to the following values of the chain indices:
$0,a:m=25$ (solid line),
$0,a:m=50$ (long--dashed line), $a:n=25,a:m=50$ (short--dashed line)
and $a:n=50,b:m=50$ (dotted line). Note also that these curves run 
from bottom to top in the left of the figure.
}
\end{figure}

Clearly, an open chain may be viewed as a star with $f=2$ arms,
so one would like to understand the effect of increasing the arms number,
which will be studied in Sec. \ref{sec:scaling}, but we shall
first analyse different correlation functions in a star with
sufficiently large $f$.
Thus, in Fig. \ref{fig:gkStar} we present the functions $\hat{g}^{(2)}_{ij}$
for different $ij$ for a star with $f=12$ arms in the coil state.

Star architecture allows us to use the arms symmetry to increase
averaging statistics. Therefore, it is sufficient to consider
correlations of three types: core--arm, intra--arm and inter--arm.
As can be seen from Fig. \ref{fig:gkStar}, the shapes of functions 
$\hat{g}^{(2)}_{ij}$ corresponding to these three types are very
different indeed. 

The core monomer plays a very special r\^{o}le, so the solid
(core--half arm) and long--dashed (core--end monomer)
curves in Fig. \ref{fig:gkStar} show pronounced correlation holes
at small $\hat{r}$ due to stronger steric repulsion from
the core monomer as $f$ increases. Interestingly, $\hat{g}^{(2)}_{ij}$
grows faster at small $\hat{r}$ in the core--end case than in
the core--half case. This may be explained by entropic reasons
as it is easier for an end monomer to form a single long loop and 
return towards the core.

The intra--arm correlation function drawn
as short--dashed curve in Fig. \ref{fig:gkStar} has a shape and meaning
similar to that of the corresponding function in an open chain 
(short--dashed curve in Fig. \ref{fig:gkOpen}) since the core monomer
and the rest of the star are typically far from the part of the arm 
in question.
Analogously, the inter--arm end--end correlation function
(dotted curve in Fig. \ref{fig:gkStar}) has a similar shape and
meaning to that of the end--end function in an open chain
(solid curve in Fig. \ref{fig:gkOpen}), although the correlation
hole is smaller in the latter case. This is because there are more
possibilities for any two ends in a star to encounter each other
than to do so for the only two ends of an open chain. This effect will be 
numerically investigated for different $f$ in Sec. \ref{sec:scaling}.
Thus, correlation functions of a star are essentially $ij$-dependent
even within monomers of identical topological types (i.e. inner, end
or core).

\section{Scaling relations}\label{sec:scaling}

\subsection{Open and ring chains}

In this section we shall quantify the behaviour of rescaled
correlation functions in a good solvent.
According to Refs. \onlinecite{CloizeauxBook,Schafer} these, at least
in the well known case of an open flexible chain, 
can be described as a power law times a stretched
exponential, so that
we can fit $\hat{g}^{(2)}_{ij}(\hat{r})$ there via the following function,
\begin{equation} \label{FitA}
\hat{g}^{(2)}_{ij}(\hat{r}) = A_{ij}\,\hat{r}^{\theta_{ij}}\,
\exp\left( -B_{ij}\,\hat{r}^{\delta_{ij}} \right).
\end{equation}
For brevity we shall suppress the $ij$ indices bellow.
Due to the two normalisation conditions in Eq. (\ref{normtwo}) constants
$A$ and $B$ can be immediately calculated and expressed via 
$\theta$ and $\delta$: 
\begin{equation}
B=\frac{\Gamma((5+\theta)/\delta)}{\Gamma((3+\theta)/\delta)^{\delta/2}}, \qquad
A=\frac{\delta\,B^{(3+\theta)/\delta}}{4\pi\,\Gamma((3+\theta)/\delta)}.
\end{equation}

The exponents $\delta$ and $\theta$, which in the case of 
end--end correlations of an open chain
is denoted as $\theta_0$, can be expressed via,
\begin{equation}\label{DeCl}
\delta=\frac{1}{1-\nu}, \qquad \theta_0=\frac{\gamma-1}{\nu},
\end{equation}
where $\nu$ has the meaning of the inverse fractal dimension of the system
and $\gamma$ is related to the number of different polymer conformations
\cite{CloizeauxBook,Schafer}.

We should, strictly speaking, comment at this stage that for small
distances the $\theta_0$ exponent should, in fact, be replaced by
another value $\ae=(1-\gamma+\nu d -d/2)/(1-\nu)=0.249\pm 0.011$, but
because $\ae$ is numerically very close to $\theta_0$ in $d=3$ dimensions,
one can use Eqs. (\ref{FitA},\ref{DeCl}) in a single form for all distances.
Such formula is often called the des Cloizeaux scaling form
\cite{CloizeauxBook,Schafer}. This expression also works
very well for other types of correlations and we shall fit our
data via Eq. (\ref{FitA}) below.

The exponents $\delta$ and $\nu$ 
have been derived in Refs. \onlinecite{Guillou,Brezin}.
The theoretical
values that we shall quote in this paper have been updated by a
higher order calculations in Ref. \onlinecite{Guida},
\begin{equation}
\gamma=1.1596 \pm 0.0020, \qquad \nu=0.5882 \pm 0.0011.
\end{equation}

The exponents $\theta_{\upsilon}$, $\upsilon=0,1,2$ 
related to the probabilities of contacts of the end--end, end--internal
and internal--internal monomers in an open chain respectively,
have been calculated by des Cloizeaux
in Ref. \onlinecite{desCloizeaux80} by the renormalisation group technique
based on $\varepsilon=4-d$ expansion
for the Euclidean field theory, as well as by Duplantier 
in Ref. \onlinecite{DuplantierLet} and Oono {\it et al} in Ref.
\onlinecite{Oono} based on the Edwards' model.

In Tab. \ref{tab:1} we present the exponents, based on fitting of our 
Monte Carlo data, as well as the best theoretical values known to us
(the latter and the corresponding Monte Carlo values under comparison
are set in bold face).
Fitting has been done via the the nonlinear least--squares
(NLLS) Marquardt--Levenberg method \cite{NumerRecip}
by means of the {\tt fit} function in the {\tt gnuplot} software ver. 3.7.0.
{\tt Fit} reports parameter error estimates which are
obtained from the variance--covariance matrix after the final iteration.
By convention, these estimates are called `standard errors' and they
have been reported in the tables.

Two types of fittings have been independently carried out. In one case,
both $\delta$ and $\theta$ have been fitted, while in the other case,
$\delta$ was fixed at the theoretical value and hence a more accurate
estimate has been achieved. The latter exponent will be denoted as
$\theta_{\delta\,fix}$.

One should bear in mind the reservations that universal exponents values 
are only asymptotically approached as $N$ increases, and that 
assuming the same power law for both small and large distances
is only approximate.
Yet, we find our results in a good agreement with the best 
of theoretical estimates.

First of all, the exponent $\delta$ in cases when fitting errors are the
smallest seems quite close to $\delta_{theor}$. The agreement is particularly
good in case ${\bf 1'}$, where $\theta_{\delta\,fix}$ is also 
very close to theoretical value $\theta_1$.
This may be in part due to the fact that here the two internal monomers
are well separated within the chain, reducing finite--size behaviour.
It is interesting that $\delta$ for the ring is also close to
the current exponent,
which indicates that as $N$ increases the same universal exponent
value should be achieved in this case also. 
We should comment that a somewhat
longer ring with $N=300$ is equivalent to only a $N=150$ open chain though.
As for the end--end case ${\bf 0}$, it is somewhat more intrinsically
noisy, since ends have more entropic freedom, so that the agreement
with the much more accurately known $\theta_0$ is only satisfactory.

In the internal--internal case ${\bf 2}$ our fitting error is
rather minimal, thus our value of $\theta_{\delta\,fix}$ here is
also very consistent with the ten times more accurate result for the ring.
Such improved accuracy is achieved for the ring because here one can
average over all pairs of monomers separated by $n=|i-j|$ links
due to the kinematic symmetry described above. Thus, the error
is reduced by an order of $\sqrt{100}$, which is also seen in the data.
It is only natural that any two well separated internal monomers
in an infinitely long open chain should behave as those of
an infinitely long equivalent ring since an extra
link between the two ends should not matter.
However, the theoretical estimate for $\theta_2$ is the least accurate of all,
so that higher order calculations akin to those done for $\theta_0$ 
recently may be required here too.

Moreover, here we can see a cross--over effect discussed in Ref. 
\onlinecite{Baumgartner} at work.
By taking one monomer at the end, and by moving another from
$1/4$-chain (case ${\bf 1''}$) to $1/2$-chain (case ${\bf 1}$)
and to $3/4$-chain (case ${\bf 1'}$), the exponent $\theta_{\delta\,fix}$
would steadily decrease, and as we reach the other end (case ${\bf 0}$),
the smallest value $\theta_0$ would be reached. The tilde in front of the
theoretical exponents in Tab. \ref{tab:1} indicates that those exponents
are in the cross--over area.

\subsection{Stars}

Results from fitting of various correlations functions 
for stars with different $f$ in the manner described above 
are collected in Tab. \ref{tab:2}. In this case,
no theoretical exponents are known to us as universality
is not a feature of this system. Yet, it is instructive to
attempt the fitting even though we realise the finite--size and cross--over
nature of the resulting numbers. This should not be very off--putting
though as actually synthesized stars are indeed objects of a few arms of
very finite length at present.

First of all, we find again that the stretching exponent $\delta$ agrees
within fitting errors with the theoretical value for open and ring chains
$\delta_{theor}$. Thus, we can concentrate on the discussion of a more accurate
exponent, $\theta_{\delta\,fix}$. Clearly, the core monomer has a very special
r\^{o}le, and as the number of arms grows,
the steric repulsion around it becomes really significant.
This leads to a rapid expansion of the correlation hole
as $f$ changes from 3 to 12, so that values of $\theta_{\delta\,fix}$ grow
rapidly with $f$ for the correlations functions of the core with the middle and
ends of arms (two first rows). As for the intra--arm correlations generally,
these behave more and more as in an open chain as long as the monomers in
question are away from the core. So the middle--end intra--arm exponents
(third row)
$\theta_{\delta\,fix}$ are fairly independent of $f$ and are close in
numerical values to the corresponding exponent $\theta_1$ of an open chain.

Now let us discuss the inter--arm exponents. 
The middle--middle (fourth row) and end--end (last row) inter--arm
exponents $\theta_{\delta\,fix}$ start from values close to those of the
corresponding exponents of an open chain $\theta_2$ and $\theta_0$ respectively
at $f=3$, but both of them decrease significantly as number of arms $f$
increases. Finally, the end--middle inter--arm exponent $\theta_{\delta\,fix}$
(fifth row) starts from a value fairly close to $\theta_1$ and decreases
only slowly with increasing $f$.

\section{Asphericity characteristics}\label{sec:aspher}

In this section we would be interested to investigate the average shape of
conformations, for which we have studied the spherically
averaged pair correlation functions above.

The shape tensor in Eq. (\ref{tens}) is merely a slight generalisation
of its 3-d trace ${\cal R}_g^2$ in Eq. (\ref{Rg}). However, since the space
rotational isotropy is enforced on average, one way of obtaining a
non--trivial spatial information is to diagonalise the matrix 
$Q^{\alpha\beta}$ for each member of the statistical ensemble
and then to average some useful observables constructed out of its
eigenvalues. Eq. (\ref{La}) introduces the averages of normalised 
sorted eigenvalues (so that $\lambda^{(1)} \ge \lambda^{(2)} \ge 
\lambda^{(3)}$), 
whereas Eqs. (\ref{Aa}) and (\ref{Ss}) introduce
the asphericity index ${\cal A}_3$ and the ellipsoid index ${\cal S}_3$.
The latter allows one to further distinguish a prolate ellipsoid (rod--like)
from an oblate one (pancake--like). Finally, variables $\hat{{\cal A}}_3$ and
$\hat{{\cal S}}_3$ in Eqs. (\ref{hatAS}) introduce related fractions of 
averages, which although less informative are easier to compute analytically
\cite{Schafer}.

In Tab. \ref{tab:3} we have collected these asphericity characteristics
for the systems under study. First of all, we may observe that the
`ellipsoid aspect ratios' $\lambda^{(a)}$ reflect the shape fairly
well, although the asphericity index ${\cal A}_3$ and the ellipsoid
index ${\cal S}_3$ express it in an even more clear way. As for the
observables $\hat{{\cal A}}_3$ and $\hat{{\cal S}}_3$, they follow
the trends of ${\cal A}_3$ and ${\cal S}_3$ respectively, although in
a somewhat less sensitive manner.

Now then, for the ring, open coil and the globule
we see that the values depend little on $N$, if the system size
is large enough, approaching their universal limits.
The coil of a ring (first row) has a smaller asphericity and 
prolateness (rod--like shape) than the corresponding values for an
open chain (third row).
We may note also that our values for an open coil are fairly close
to those reported in Refs. \onlinecite{Jagodzinski,Schafer}.
The semi--stiff ring (fourth row) becomes, clearly, more
aspherical and prolate as stiffness $\lambda$ increases
at fixed $N$. If, however, $\lambda$ is kept fixed,  and $N$ increases
both ${\cal A}_3$ and ${\cal S}_3$ decrease approaching
the flexible chain limit since the ratio $\lambda/N$ presents 
the relevant degree of flexibility of the polymer.

The globule of a ring (second row) is nearly perfectly
spherical and only very slightly prolate, as should be expected,
and these asymmetries diminish even further as $N$ increases. 
However, at the same time, the non--equality of $\lambda^{(a)}$ is 
more noticeable here, which shows that they provide less 
informative asphericity observables.

Finally, stars (fifth row) starting from a fairly non--spherical
and prolate shape at $f=3$ become significantly
more spherical and less prolate as the number of arms $f$ increases.
This, of course, is consistent with the intuitive idea that a star
with a very large number of relatively short 
arms behaves as a sphere due to strong steric repulsions.

\section{Static structure factors}

The static structure factor (SSF) is of interest due to its relation to the
light and neutron scattering techniques and it is standard to
introduce it in the rescaled Kratky form via, 
\begin{equation}
\label{ssfResc}
\hat{S}(\hat{p}) = \frac{p^2\,{\cal R}_g^2}{N}\,S(p), \qquad
\hat{p} = p\,\sqrt{{\cal R}_g^2},
\end{equation}
which has an advantage of being less sensitive on $N$.

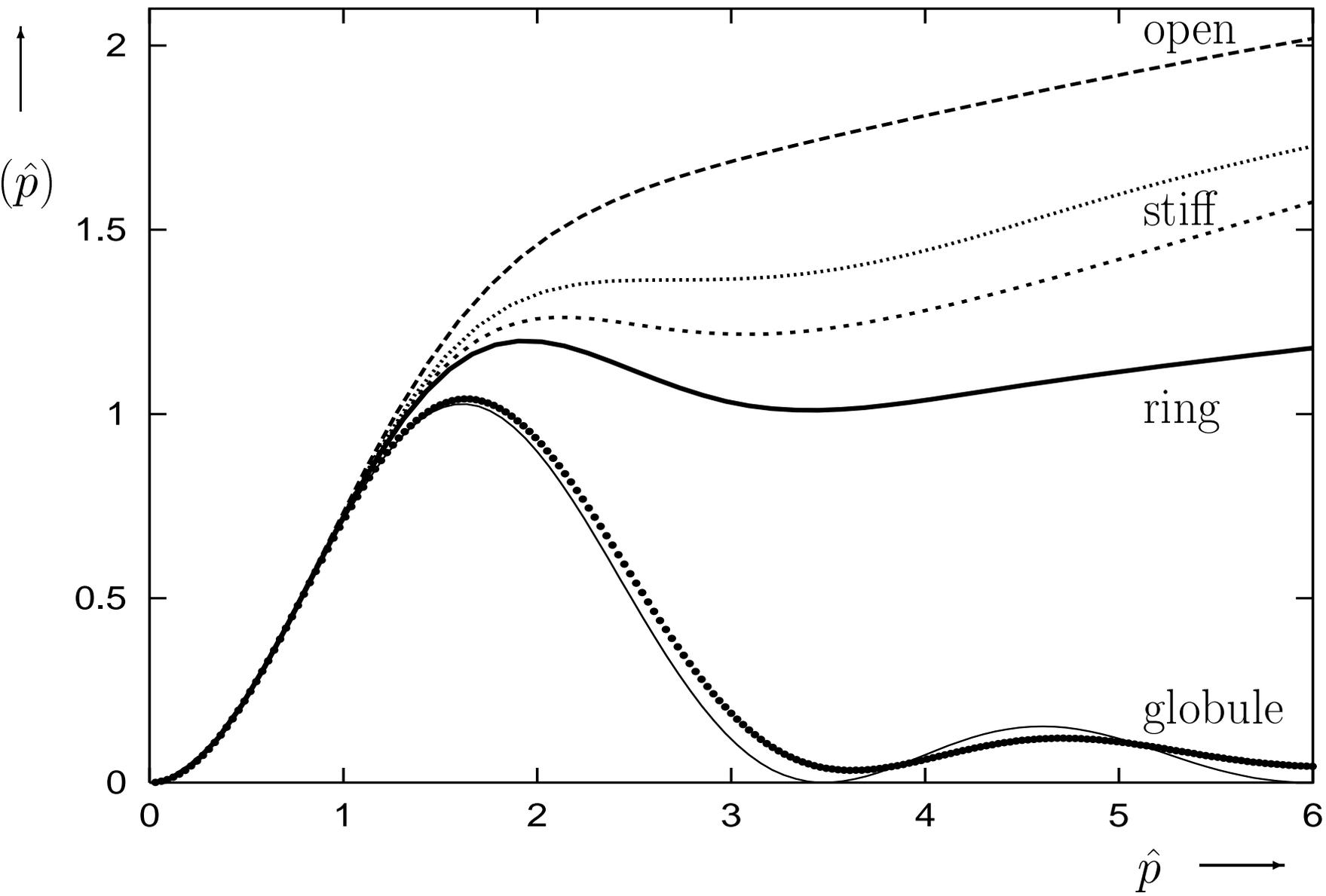
\begin{figure}
\caption{ \label{fig:SkAll}
Plot of the rescaled static structure factor $\hat{S}(\hat{p})$ 
of homopolymers vs the 
rescaled wave number $\hat{p}$.
Curves correspond to the following chains from top to bottom:
open flexible coil with $N=200$ (long--dashed line),
semi--flexible ($\lambda=1$) ring coil with $N=200$ (dotted line),
semi--flexible ($\lambda=1$) ring coil with $N=100$ (short--dashed line),
flexible ring coil with $N=200$ (thick solid line), 
globule ($V_0=6\,k_B T$) of flexible ring with $N=200$ (solid circles curve),
and the solid sphere with radius $R_s=\sqrt{(5/3){\cal R}_g^2}$ 
(thin solid line).
}
\end{figure}

The factor $p^2$ allows one to emphasise the large $p$ behaviour more
clearly. Indeed, it is well known \cite{Schmitz}
that $S(p)\ \propto p^{-1/\nu}$
with $\nu=3/5$ and $1/2$ for the repulsive and ideal coils
respectively. Thus, $\hat{S}(\hat{p})$ would be increasing,
reaching a constant asymptote, or decreasing for the repulsive coil,
ideal coil and globule respectively.
This is indeed the case in Fig. \ref{fig:SkAll}, where we depict 
$\hat{S}(\hat{p})$ vs $\hat{p}$ for various systems. Since
curves for the coils (or the globule)
for different $N$ coincided with each other so well due to
a nearly perfect universality, we only have drawn 
one particular size in each case.

Clearly, for small $\hat{p}$ we have the identical behaviour\cite{Schmitz},
\begin{equation}
\label{SSFas}
\hat{S}(\hat{p})\simeq \hat{p}^2(1 - \hat{p}^2/3 + \ldots),
\end{equation}
in terms of the rescaled variables in all cases. The SSF
of the globule has a characteristic oscillating behaviour, which shows
its dense structure. 
It is instructive to compare this with the SSF of a solid sphere\cite{Schmitz}
of a radius $R_s$,
\begin{equation}
\label{SSFspher}
S(p)=\left(\frac{3}{(R_s p)^3}\biggl(\sin(R_s p)-(R_s p) \cos (R_s p)\biggr)\right)^2.
\end{equation}
By requiring that the cumulant law Eq.\ (\ref{SSFas}) 
is universal in terms of the rescaled variables one determines
the radius to be $R_s=\sqrt{(5/3){\cal R}_g^2}$.
One can see in Fig. \ref{fig:SkAll} that the SSF of the globule
is in a perfect agreement with that of the solid sphere up to the
peak point. At larger $\hat{p}$ both functions have rather similar
shape, however the SSF of the globule has somewhat smaller amplitude
of oscillations, which reflects fluctuations in shape and size of the globule. 
As additional analysis shows, with
increasing $N$ the minima of the rescaled SSF of the globule
decrease in value, so that $\hat{S}(\hat{p})$ asymptotically 
approaches that of the solid sphere.

The SSF $\hat{S}(\hat{p})$ of an open coil has a power--law increasing
behaviour, which is preserved for a ring coil as  well, although the latter
has a slight peak, showing a more spherical structure of a ring
in accord with Sec. \ref{sec:aspher}. 
However, when we introduce even a moderate stiffness for a ring coil,
the plot of $\hat{S}(\hat{p})$ starts to significantly deviate
from that of a flexible ring, and also becomes quite $N$
dependent. This results from the loss of master function universality
in the rescaled correlation functions in terms of $N$, which we have 
seen for a semi--flexible coil in subsec. \ref{subsec:stiff}.

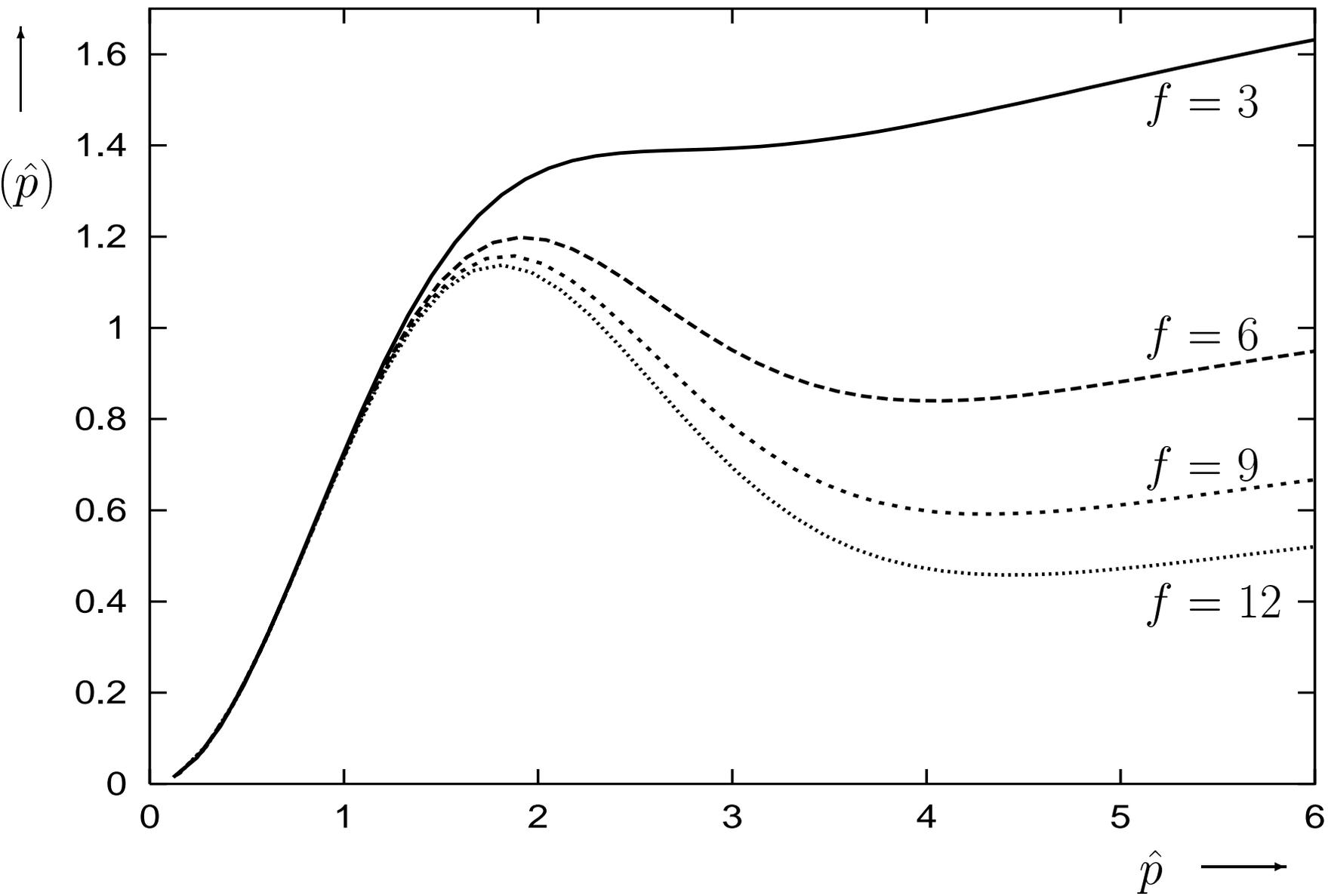
\begin{figure}
\caption{ \label{fig:SkStar}
Plot of the rescaled static structure factor $\hat{S}(\hat{p})$ 
of star homopolymers vs the 
rescaled wave number $\hat{p}$.
Curves correspond to stars with the following numbers of 
arms from top to bottom:
$f=3$ (solid line), $f=6$ (long--dashed line), 
$f=9$ (short--dashed line), $f=12$ (dotted line).
}
\end{figure}

Now, if we turn our attention to the structure factors
of stars with different numbers of arms in Fig. \ref{fig:SkStar},
we can see how the starting case of $f=3$, which is quite
reminiscent of the curve of an open chain, transforms into a shape 
corresponding to more spherical objects as $f$ increases 
(as we have also seen in
Sec. \ref{sec:aspher}). Again, the $f$-dependence here is quite dramatic,
although $\hat{S}(\hat{p})$ still increases for large $\hat{p}$ as
the stars are in an extended state.

\section{Conclusion}\label{sec:concl}

In this paper we have studied the correlation functions and
related observables for a single homopolymer chain
with several macromolecular architectures.

A ring case is the most symmetric one and here
we have discovered the universality of
the dimensionless correlation functions $\hat{g}^{(2)}_n$
for all chain index separations beyond small scale $n^{\dagger} \simeq 5$.
Here the universality also holds for different chain lengths $N$.

The universality appears to be valid in terms of $n$ for the globular
state as well, although the scale index is of order 
$n^{*}\propto N^{2/3}$ there, and
the correlation functions have a complicated oscillating behaviour.
Interestingly, the original function $g_n^{(2)}(r)$ is also
$n$-universal for the globule, which is consistent with the behaviour of 
the pair correlation function akin to that of simple liquids, 
so that the peaks in $g_n^{(2)}(r)$ occur at integer multiples 
of the hard core diameter $r=m\,d$, requiring for ${\cal D}_n$ 
to become constant beyond $n^{*}$.

For the coil the functions $\hat{g}^{(2)}_n$ can be accurately
fitted via a power law times a stretched exponential --- the des Cloizeaux
scaling function Eq. (\ref{FitA}). The stretching exponent $\delta$,
related to the fractal dimension of the chain, as well as the
power exponent $\theta$, related to the correlation hole effect,
are found to be close to the results of theoretical calculations
in the special case when the Kuhn length is chosen equal to the
hard sphere diameter: $\ell/(\sqrt{\kappa}d)=1$. This strong repulsion
regime is a special one because it most closely corresponds to the continuous
model of the polymer chain.

However, we have found that decreasing $\kappa$
results in a significant change in the power law exponent $\theta$,
even though the long--distance exponent is less affected by that.
This may be easily understood as a cross--over to the Gaussian chain
behaviour, namely $\theta\to 0$ and $\delta\to 2$ as 
$\ell/(\sqrt{\kappa}d)\to\infty$. Strong sensitivity of the correlation
hole on the model parameters also demonstrates that care should be taken
when comparing universal renormalisation group predictions with results from
simulations of concrete systems.

Clearly, the ends play a special r\^{o}le for an open chain.
Therefore, one has to distinguish the end--end, end--internal
and internal--internal correlations with the corresponding exponents
$\theta_{\upsilon}$ respectively. The exponent $\delta$ is found
to be universal for strongly--repulsive flexible coils of any architecture 
and numerically very close to that of 
the ring chain and to the theoretical prediction. Fitting 
$\hat{g}^{(2)}_{ij}(\hat{r})$ via the des Cloizeaux
expression in the strong repulsion regime ($\ell/(\sqrt{\kappa}d)=1$)
has produced values fairly close to the results of theoretical
calculations. The effect of cross--over, when the internal monomer is
placed in different locations along the chain, is also observed.

In our study of the semi--flexible ring chain we have observed a total
lack of the $n$- or $N$- universality in $\hat{g}^{(2)}_n(\hat{r})$.
The correlation hole at small $\hat{r}$ is particularly sensitive on $n$
even for chain separations $n$ well exceeding the persistence
length parameter $\propto \lambda$. Naturally, the correlation hole
expands for small $n$ due to the stiffness of the chain
and consequently contracts for larger $n$. We may note also that 
in the latter case the
behaviour of $\hat{g}^{(2)}_{n}(\hat{r})$ is somewhat similar to that of
the chain with loose springs, leading to smaller values of $\theta$.

In the case of star polymers one has to distinguish the special
r\^{o}le of the core monomer in addition to the end monomers, as
well as the intra-- and inter--arm correlations. Generally,
in the coil state here we have a des Cloizeaux scaling behaviour with
nearly universal $\delta$, but with the exponent $\theta$ being strongly
sensitive on the monomer indices $ij$, number of arms $f$, as well
as on the arm length $(N-1)/f$. A somewhat more stable behaviour
for $\theta$ was only seen for the intra--arm correlations
when both monomers are fairly far along the arm from the core monomer,
agreeing with the numerical values of $\theta_{\upsilon=1,2}$ in an open
chain.

We have also studied the average shape of the above described polymers
by means of computing the asphericity characteristics.
We have found that the coil of an open chain
is fairly aspherical and prolate (rod--like),
while the coil of a ring is less so. Interestingly, the coil of
a semi--stiff ring becomes more aspherical and prolate than that of the flexible
ring as $\lambda$ increases or $N$ decreases.
Star polymers in the coil state are fairly aspherical and prolate
for small numbers of arms $f$, becoming more spherical as $f$ increases.
Finally, the globule of a ring is nearly perfectly spherical.

The static structure factor, related to the sum of the 3-d Fourier
transforms of the correlation functions, 
has been investigated in terms of its appropriate scaling variables.
Behaviour at small rescaled wave numbers $\hat{p}$ is universal,
while the large $\hat{p}$ tail scales as a power law with the exponent 
related to the fractal dimension of the object.
In cases of the flexible open chain, the ring coil and the globule the
functions $\hat{S}(\hat{p})$ have shapes practically independent
of polymer length $N$, which however is not true for semi--flexible
polymers. Likewise, functions $\hat{S}(\hat{p})$ are strongly
dependent on the number of arms $f$ for star polymers.

Finally, let us emphasise that we have only discussed the globule
of a ring here because most of the results would be little, if at all,
affected by the change of the connectivity matrix in the dense
globular state, although this is a much more complicated issue
meriting a separate study in case of fairy stiff chains,
where, as we know, transitions into toroidal globules
take place \cite{Torus}.

Thus, in this paper, we have seen  that the degree of universality of
various scaling functions and exponents is subject to more
serious limitations than have been previously realised.
The knowledge of various correlation functions of the polymer
chain and their response to changes of various parameters provides
us with the most detailed insight into the structure of the system.
Development of theoretical methods, which could reproduce
various features observed in this paper, is important
for tackling larger system sizes and more complicated
systems such as heteropolymers, for which simulations have severe
limitations. This work is currently in progress and we hope to be able
to return to such subject in the near future.

\acknowledgments

The authors are grateful for interesting discussions to
Professors~F.~Ganazzoli, T.~Garel, J.~Zinn-Justin, M.~Daoud,
B.~D\"{u}nweg,
and support from Enterprise Ireland: basic research grant SC/99/186
and international collaboration grants IC/2001/074 and BC/2001/034.



\newpage
\section*{Figure Captions}

\newpage

\section*{Tables}

\begin{table}
\caption{\label{tab:1}
Comparison of the exponents $\delta$ and $\theta$ between the results from
Monte Carlo simulations and theoretical results (these are supplied with
subscript $theor$) for ring and open homopolymer coils.  Values in the
second and fourth columns have been obtained by a four--parametric
fit via Eq. (\ref{FitA}) and the values in the fifth column (denoted
as $\theta_{\delta fix}$ have been obtained by a three--parametric fit
via the same equation, but with fixed $\delta=\delta_{theor}\equiv 2.428363$.
The last column contains references to works with theoretical exponents.
The notations in the first column follow des Cloizeaux convention:
{\bf 0} --- end--end monomers, {\bf 1} ---
end--middle, {\bf 1}' --- end--three quarters, {\bf 1}'' --- 
end--one quarter, and {\bf 2} --- one quarter--three quarters of the 
chain respectively. Here and below reported errors are those 
from the fitting procedure only and do not necessarily
account for statistical and other simulation errors.
}
\vskip 5mm
\begin{tabular}{|l|l|l|l|l|l|l|}
 & $\delta$ & $\delta_{theor}$ & $\theta$ & $\theta_{\delta fix}$ & $\theta_{theor}$ & Ref. \\ \hline
Open $N=200$\  {\bf 0} & $2.11 \pm 0.07$ & $2.428 \pm 0.001$ & $0.36 \pm 0.02$ & ${\bf 0.30 \pm 0.02}$ & ${\bf 0.271 \pm 0.002}$ & \onlinecite{Guida} \\ \hline  

Open $N=200$\  {\bf 1} & $2.23  \pm 0.04 $ & $2.428 \pm 0.001$ & $0.56 \pm 0.01$ & $0.51 \pm 0.01$ & $\sim 0.46    $ & \onlinecite{CloizeauxBook} \\ \hline  

Open $N=200$\ {\bf 1'} & ${\bf 2.42  \pm 0.04} $ & ${\bf 2.428 \pm 0.001}$ & $0.45 \pm 0.01$ & ${\bf 0.45  \pm 0.01}$ & ${\bf 0.459 \pm  0.003}  $ & \onlinecite{CloizeauxBook} \\ \hline  

Open $N=200$\ {\bf 1''} & $2.04 \pm 0.08 $ & $2.428 \pm 0.001$ & $0.68 \pm 0.03$ & $0.56 \pm 0.02$ & $\sim 0.46    $ & \onlinecite{CloizeauxBook} \\ \hline  

Open $N=200$\ {\bf 2} & $2.39  \pm 0.07 $ & $2.428 \pm 0.001$ & $0.81 \pm 0.02$ & ${\bf 0.80 \pm 0.01}$ & ${\bf 0.71  \pm 0.05}  $ & \onlinecite{CloizeauxBook} \\ \hline\hline  

Ring $N=300$ & ${\bf 2.46 \pm 0.01}$ & ${\bf 2.428 \pm 0.001}$ & $0.79 \pm 0.006$ & ${\bf 0.816  \pm 0.002}$ &  &  \\ 
\end{tabular}
\end{table}

\begin{table}
\caption{\label{tab:2}
Values of the exponents $\delta$ and $\theta$ for star polymers with $f=3,6,12$
arms and $(N-1)/f=50$ arm length in a good solvent.
These have been obtained by a four--parametric
fit via Eq. (\ref{FitA}) and the values for 
$\theta_{\delta fix}$ have been obtained by a three--parametric fit
via the same equation, but with fixed $\delta=\delta_{theor}$.
The following notations for monomer pairs have been adopted:
$a:n,b:m$, where $a,b$ number arms and $n,m$ number monomers within arms
and $0$ refers to the core monomer.
}
\vskip 5mm
\begin{tabular}{|l|l|l|l|l|}
 & & $f=3$ & $f=6$ & $f=12$ \\ \hline
           & $\delta$     & $2.86 \pm 0.07$ & $2.66 \pm 0.04$ & $2.66 \pm 0.03$ \\
$0,a:m=25$ & $\theta$     & $1.07 \pm 0.03$ & $2.01 \pm 0.03$ & $3.03 \pm 0.04$ \\
 & $\theta_{\delta\ fix}$ & $1.25 \pm 0.02$ & $2.19 \pm 0.02$ & $3.30 \pm 0.02$ \\ \hline
 
           & $\delta$     & $2.54 \pm 0.06$ & $2.57 \pm 0.04$ & $2.55 \pm 0.03$ \\
$0,a:m=50$ & $\theta$     & $0.66 \pm 0.02$ & $1.05 \pm 0.02$ & $1.58 \pm 0.02$ \\
 & $\theta_{\delta\ fix}$ & $0.70 \pm 0.01$ & $1.12 \pm 0.01$ & $1.66 \pm 0.01$ \\ \hline

           & $\delta$     & $2.31 \pm 0.06$ & $2.55 \pm 0.05$ & $2.39 \pm 0.03$ \\
$a:n=25,a:m=50$ & $\theta$     & $0.57 \pm 0.02$ & $0.47 \pm 0.02$ & $0.54 \pm 0.01$ \\
 & $\theta_{\delta\ fix}$ & $0.54 \pm 0.01$ & $0.51 \pm 0.01$ & $0.53 \pm 0.01$ \\ \hline

           & $\delta$     & $2.34 \pm 0.06$ & $2.34 \pm 0.03$ & $2.09 \pm 0.05$ \\
$a:n=25,b:m=25$ & $\theta$     & $0.84 \pm 0.03$ & $0.72 \pm 0.01$ & $0.59 \pm 0.03$ \\
 & $\theta_{\delta\ fix}$ & $0.81 \pm 0.01$ & $0.69 \pm 0.01$ & $0.52 \pm 0.01$ \\ \hline

           & $\delta$     & $2.50 \pm 0.05$ & $2.47 \pm 0.03$ & $2.37 \pm 0.03$ \\
$a:n=25,b:m=50$ & $\theta$     & $0.41 \pm 0.01$ & $0.43 \pm 0.01$ & $0.36 \pm 0.01$ \\
 & $\theta_{\delta\ fix}$ & $0.43 \pm 0.01$ & $0.44 \pm 0.01$ & $0.35 \pm 0.01$ \\ \hline

           & $\delta$     & $2.25 \pm 0.08$ & $2.43 \pm 0.04$ & $2.29 \pm 0.03$ \\
$a:n=50,b:m=50$ & $\theta$     & $0.35 \pm 0.03$ & $0.25 \pm 0.01$ & $0.22 \pm 0.01$ \\
 & $\theta_{\delta\ fix}$ & $0.31 \pm 0.01$ & $0.26 \pm 0.01$ & $0.18 \pm 0.01$ \\ 
\end{tabular}
\end{table}

\begin{table}
\caption{\label{tab:3}
Asphericity characteristics $\lambda^{(a)}$ (Tab. a), 
${\cal A}_3$, ${\cal S}_3$,
and $\hat{{\cal A}}_3$, $\hat{{\cal S}}_3$ (Tab. b)
defined by Eqs. (\ref{La},\ref{Aa},\ref{Ss},\ref{hatAS}) respectively
from Monte Carlo simulations. Here notations for chains correspond to:
{\bf ring} --- flexible ring homopolymer coils, 
{\bf globule} --- flexible homopolymer rings in a poor solvent 
($V_0=6\,k_B T$), {\bf open} --- flexible open homopolymer coils, 
{\bf stiff} --- semi--flexible ring homopolymer 
coils with stiffness constants $\lambda=1,5$, and
{\bf star} --- flexible star homopolymers in a good solvent with 
arm length $(N-1)/f=50$. 
}

\newpage
\centerline{{\bf a}}
\vskip 3mm

\begin{tabular}{|l|ccc|}
System & $\lambda^{(1)}$ & $\lambda^{(2)}$ & $\lambda^{(3)}$ 
\\
\hline
{\bf ring} &     &        &                \\
$N=100$ &  $0.637 \pm 0.001$   &  $0.2637\pm 0.0008$  &  $0.0995\pm 0.0004$   \\
$N=200$ &  $0.637 \pm 0.001$   &  $0.2643\pm 0.0008$  &  $0.0986\pm 0.0004$  \\
$N=300$ &  $0.639 \pm 0.001$   &  $0.2651\pm 0.0008$  &  $0.0957\pm 0.0004$  \\
$N=500$ &  $0.640 \pm 0.001$   &  $0.2654\pm 0.0008$  &  $0.0945\pm 0.0004$  \\
\hline
{\bf globule} &  &        &        \\
$N=100$ &  $0.4230\pm 0.0004$  &  $0.3242\pm 0.0003$  &  $0.2529\pm 0.0001$  \\
$N=150$ &  $0.4098\pm 0.0002$  &  $0.3265\pm 0.0002$  &  $0.2637\pm 0.0001$  \\
$N=200$ &  $0.4021\pm 0.0002$  &  $0.3277\pm 0.0002$  &  $0.2702\pm 0.0001$  \\
\hline
{\bf open} &     &        &      \\
$N=150$ &  $0.750 \pm 0.0015$  &  $0.185 \pm 0.001 $  &  $0.0653\pm 0.0004$  \\
$N=200$ &  $0.7513\pm 0.0007$  &  $0.1835\pm 0.0005$  &  $0.0652\pm 0.0002$   \\
\hline
{\bf stiff} $\lambda = 1$ &    &   &    \\
$N=100$ &  $0.6529\pm 0.0006$  &  $0.2536\pm 0.0005$  &  $0.0935\pm 0.0003$  \\
$N=200$ &  $0.6445\pm 0.0006$  &  $0.2562\pm 0.0005$  &  $0.0993\pm 0.0003$  \\
$N=300$ &  $0.6418\pm 0.0007$  &  $0.2571\pm 0.0005$  &  $0.1011\pm 0.0003$  \\
{\bf stiff} $\lambda = 5$ &    &   & \\
$N=100$ &  $0.6905\pm 0.0006$  &  $0.2447\pm 0.0005$  &  $0.0648\pm 0.0003$  \\
$N=200$ &  $0.6712\pm 0.0006$  &  $0.2473\pm 0.0005$  &  $0.0815\pm 0.0003$  \\
$N=300$ &  $0.662 \pm 0.001 $  &  $0.2537\pm 0.0005$  &  $0.0841\pm 0.0003$  \\
\hline
{\bf star} &     &        &        \\
$f=3$   &  $0.6752\pm 0.0009$  &  $0.2451\pm 0.0007$  &  $0.0797\pm 0.0004$  \\
$f=6$   &  $0.5492\pm 0.0007$  &  $0.3036\pm 0.0005$  &  $0.1472\pm 0.0005$  \\
$f=9$   &  $0.4990\pm 0.0006$  &  $0.3163\pm 0.0004$  &  $0.1847\pm 0.0004$   \\
$f=12$  &  $0.4741\pm 0.0006$  &  $0.3194\pm 0.0003$  &  $0.2065\pm 0.0004$  \\
\end{tabular}

\newpage

\centerline{{\bf b}}
\vskip 3mm

\begin{tabular}{|l|cc|cc|}
System & ${\cal A}_3$ & ${\cal S}_3$ & $\hat{{\cal A}}_3$ & $\hat{{\cal S}}_3$ \\
\hline
{\bf ring}         &        &        &        &        \\
$N=100$ &  
$0.253 \pm 0.0015$ & $0.185 \pm 0.005 $ & $0.282 \pm 0.004 $ & $0.243 \pm 0.008$ \\
$N=200$ &
$0.254 \pm 0.0016$ & $0.184 \pm 0.005 $ & $0.283 \pm 0.004 $ & $0.232 \pm 0.008$ \\
$N=300$ &  
$0.258 \pm 0.0014$ & $0.187 \pm 0.005 $ & $0.286 \pm 0.004 $ & $0.236 \pm 0.008$ \\
$N=500$ &  
$0.260 \pm 0.0015$ & $0.188 \pm 0.005 $ & $0.287 \pm 0.004 $ & $0.234 \pm 0.008$ \\
\hline
{\bf globule}         &        &        &        &        \\
$N=100$ &  
$0.0255\pm 0.0001$ & $0.00317\pm 0.00007$ & $0.0261\pm 0.0002$ & $0.00344\pm 0.00009$ \\
$N=150$ &  
$0.0192\pm 0.0001$ & $0.00173\pm 0.00004$ & $0.0188\pm 0.0002$ & $0.00184\pm 0.00007$ \\
$N=200$ &  
$0.0155\pm 0.0001$ & $0.00118\pm 0.00003$ & $0.0153\pm 0.0002$ & $0.00125\pm 0.00005$ \\
\hline
{\bf open}         &        &        &        &        \\
$N=150$ &  
$0.436 \pm 0.0012$ & $0.550 \pm 0.004 $ & $0.548 \pm 0.002 $ & $0.906 \pm 0.006$ \\
$N=200$ &  
$0.4297\pm 0.0006$ & $0.539 \pm 0.002 $ & $0.5492\pm 0.0008$ & $0.928 \pm 0.003 $ \\
\hline
{\bf stiff} $\lambda = 1$ &    &   &   &  \\
$N=100$ &  
$0.276 \pm 0.001 $ & $0.224 \pm 0.002 $ & $0.313 \pm 0.002 $ & $0.299 \pm 0.004$ \\
$N=200$ &  
$0.263 \pm 0.001 $ & $0.208 \pm 0.002 $ & $0.299 \pm 0.002 $ & $0.283 \pm 0.004$ \\
$N=300$ &  
$0.259 \pm 0.001 $ & $0.203 \pm 0.002 $ & $0.292 \pm 0.002 $ & $0.271 \pm 0.004$ \\
{\bf stiff} $\lambda = 5$    &   &   &   &  \\
$N=100$ & 
$0.342 \pm 0.001 $ & $0.300 \pm 0.002 $ & $0.358 \pm 0.002 $ & $0.305 \pm 0.004$ \\
$N=200$ & 
$0.307 \pm 0.001 $ & $0.263 \pm 0.002 $ & $0.336 \pm 0.002 $ & $0.304 \pm 0.004$ \\
$N=300$ & 
$0.293 \pm 0.001 $ & $0.235 \pm 0.002 $ & $0.312 \pm 0.002 $ & $0.219 \pm 0.004$ \\
\hline
{\bf star}        &        &        &        &        \\
$f=3$   &  
$0.313 \pm 0.0014$ & $0.274 \pm 0.003 $ & $0.349 \pm 0.003 $ & $0.351 \pm 0.005$ \\
$f=6$   &  
$0.140 \pm 0.001 $ & $0.0515\pm 0.0007$ & $0.145 \pm 0.002 $ & $0.057 \pm 0.001$ \\
$f=9$   &  
$0.0857\pm 0.0005$ & $0.0193\pm 0.0005$ & $0.0874\pm 0.0008$ & $0.0197\pm 0.0008$ \\
$f=12$  &  
$0.0622\pm 0.0003$ & $0.0112\pm 0.0002$ & $0.0635\pm 0.0006$ & $0.0117\pm 0.0003$ \\
\end{tabular}

\end{table}

\end{document}